\documentclass[%
 aip,
 amsmath,amssymb,
 reprint,%
]{revtex4-1}

\usepackage{graphicx}
\usepackage{dcolumn}
\usepackage{bm}
\usepackage[dvipsnames]{xcolor} 
\usepackage{amssymb} 

\usepackage[utf8]{inputenc}
\usepackage[T1]{fontenc}
\usepackage{mathptmx}
\usepackage{etoolbox}

\makeatletter
\def\@email#1#2{%
 \endgroup
 \patchcmd{\titleblock@produce}
  {\frontmatter@RRAPformat}
  {\frontmatter@RRAPformat{\produce@RRAP{*#1\href{mailto:#2}{#2}}}\frontmatter@RRAPformat}
  {}{}
}%
\makeatother
\begin{document}

\preprint{AIP/123-QED}

\title{Effect of ion irradiation on superconducting thin films}
\author{Katja Kohop\"a\"a}
\email[]{katja.kohopaa@vtt.fi}
\affiliation{QTF Centre of Excellence, VTT Technical Research Centre of Finland Ltd, P.O. Box 1000, FI-02044 VTT, Finland}
\author{Alberto Ronzani}
\affiliation{QTF Centre of Excellence, VTT Technical Research Centre of Finland Ltd, P.O. Box 1000, FI-02044 VTT, Finland}
\author{Robab Najafi Jabdaraghi}
\affiliation{VTT Technical Research Centre of Finland Ltd, P.O. Box 1000, FI-02044 VTT, Finland}
\author{Arijit Bera}
\affiliation{VTT Technical Research Centre of Finland Ltd, P.O. Box 1000, FI-02044 VTT, Finland}
\author{M\'ario Ribeiro}
\affiliation{QTF Centre of Excellence, VTT Technical Research Centre of Finland Ltd, P.O. Box 1000, FI-02044 VTT, Finland}
\author{Dibyendu Hazra}
\affiliation{QTF Centre of Excellence, VTT Technical Research Centre of Finland Ltd, P.O. Box 1000, FI-02044 VTT, Finland}
\author{Jorden Senior}
\affiliation{QTF Centre of Excellence, VTT Technical Research Centre of Finland Ltd, P.O. Box 1000, FI-02044 VTT, Finland}
\affiliation{IST Austria, Am Campus 1, 3400 Klosterneuburg, Austria}
\author{Mika Prunnila}
\affiliation{QTF Centre of Excellence, VTT Technical Research Centre of Finland Ltd, P.O. Box 1000, FI-02044 VTT, Finland}
\author{Joonas Govenius}
\affiliation{QTF Centre of Excellence, VTT Technical Research Centre of Finland Ltd, P.O. Box 1000, FI-02044 VTT, Finland}
\author{Janne S. Lehtinen}

\affiliation{QTF Centre of Excellence, VTT Technical Research Centre of Finland Ltd, P.O. Box 1000, FI-02044 VTT, Finland}
\author{Antti Kemppinen}
\affiliation{QTF Centre of Excellence, VTT Technical Research Centre of Finland Ltd, P.O. Box 1000, FI-02044 VTT, Finland}

\date{\today}

\begin{abstract}

We demonstrate ion irradiation by argon or gallium as a wafer-scale post-processing method to increase disorder in superconducting thin films. We study several widely used superconductors, both single-elements and compounds. We show that ion irradiation increases normal-state resistivity in all our films, which is expected to enable tuning their superconducting properties, for example, toward higher kinetic inductance. We observe an increase of superconducting transition temperature for Al and MoSi, and a decrease for Nb, NbN, and TiN. In MoSi, ion irradiation also improves the mixing of the two materials. We demonstrate fabrication of an amorphous and homogeneous film of MoSi with uniform thickness, which is promising, e.g., for superconducting nanowire single-photon detectors.

\end{abstract}

\maketitle

\section{Introduction} \label{introduction}

Disorder, manifested by normal-state resistivity, has significant effects on the superconducting properties of thin films, which are of continuing scientific interest~\cite{anderson_theory_1959,astafiev_coherent_2012,weitzel_sharpness_2023,mondal_phase_2011}. Ultimately, superconductivity can be observed in films up to the sheet resistance of $R_\square\sim R_Q=h/(4e^2)\approx 6.5$~k$\Omega$~\cite{jaeger_threshold_1986,sacepe_localization_2011} where $R_Q$ is the quantum resistance of Cooper pairs, $h$ is the Planck constant, and $e$ is the elementary charge. Above that, superconductivity is destroyed by fluctuations and superconductor--insulator transition arising from the localization of Cooper pairs~\cite{sacepe_localization_2011}. Below that limit, disordered superconductors are of topical applied interest: For example, high $R_\square$ yields a high sheet kinetic inductance $L_\square \simeq h R_\square/(2\pi^2\Delta)$ in the superconducting state, where  $\Delta$ is the superconducting energy gap. High $L_\square$ allows compact superinductors for quantum information processing~\cite{maleeva_circuit_2018,grunhaupt_granular_2019-1}. Disorder also enables magnetic field resilient superconductivity~\cite{samkharadze_high-kinetic-inductance_2016}, quantum phase slip (QPS) devices for quantum metrology~\cite{lehtinen_coulomb_2012, shaikhaidarov_quantized_2022}, and superconducting nanowire single-photon detectors (SNSPD) for quantum communication~\cite{goltsman_picosecond_2001, esmaeil_zadeh_superconducting_2021}. Typical $R_{\square}$ may span roughly 1--4~$\mathrm{k\Omega}$ for superinductors or QPS and 100--200~$\mathrm{\Omega}$ for SNSPDs.

High normal-state resistivity can arise either from granular or amorphous structure. However, many applications benefit from homogeneous materials where: \emph{(i)} Grain size is too small to support local superconductivity inside a single grain~\cite{gantmakher_superconductorinsulator_2010}. \emph{(ii)} Grain size is smaller than any critical dimensions of the structures made from the material. Nanostructures, in particular, become irreproducible unless the material is either amorphous or has small grains. Improving homogeneity enables uniform films also at small thicknesses which further enhances $R_\square$ and $L_\square$. In principle, irreproducibility could also be solved with novel, single-crystalline two-dimensional superconductors~\cite{zou_superconductivity_2017,cao_unconventional_2018}, but their processing currently lacks the maturity required for large-scale fabrication.

In this article, we study the usage of broad-beam ion irradiation by argon or gallium for increasing disorder in superconducting thin films. Our wafer-scale post-processing method is applicable to any material regardless of the film deposition method, but here we focus on sputtering, which typically yields polycrystalline films. Physical collisions of the ions may, e.g., cause a dense distribution of lattice defects that scatter electrons or even dismantle grains, thus improving the functional homogeneity of the materials. Since conventional superconductivity arises from the interplay between electrons and lattice, it is expected that such structural manipulation also affects the critical temperature $T_c$, but the direction of the change is less obvious. This motivates our investigation of processing several superconductors, both the most common single-element materials (Al, Nb), and compounds in two widely used groups: nitrides (TiN, NbN) and silicides (MoSi).

Already since the 1950s, several studies on single-element materials, e.g., Al, W, Ga, and Mo, have demonstrated an increase of $T_c$ when the growth of large grains has been prevented, e.g., by evaporating on a cold substrate or in the presence of oxygen, or by creating a layered structure with other materials~\cite{buckel_einflus_1954,kammerer_superconductivity_1965,abeles_enhancement_1966,strongin_effect_1967,pettit_film_1976,jaeger_threshold_1986}. Evaporating with oxygen is a key fabrication step for the modern superinductors based on granular aluminum~\cite{maleeva_circuit_2018,grunhaupt_granular_2019,grunhaupt_granular_2019-1}. In contrast, decreasing the grain size in sputtered niobium by controlling the deposition parameters yields a significant decrease of $T_c$~\cite{bose_mechanism_2005}, however proton irradiation has been observed to cause only a slight decrease of $T_c$ in Nb while increasing its upper critical field~\cite{tanatar_anisotropic_2022}. Other examples of the subtle interplay between disorder and superconductivity include, e.g., significant decrease of $T_c$ in several high-$T_c$ compounds with an $A$-15 lattice due to neutron irradiation~\cite{sweedler_atomic_1974} and the dependence of $T_c$ on the crystalline phase of tungsten~\cite{lita_tuning_2005}.

The disordered structure is unstable in many single-element materials, and already room temperature conditions are sufficient to crystallize them into relatively large grains. Disorder can be stabilized, e.g., with oxidized grain surfaces or with alloyed compounds, which may already be of interest in terms of increased $T_c$~\cite{osofsky_new_2001}. One such compound with a metastable disordered phase is MoSi, which is a well-known material for SNSPDs and can be fabricated, e.g., by sputtering from an alloy-target to a cooled substrate~\cite{bosworth_amorphous_2015} or by co-sputtering at room temperature~\cite{banerjee_characterisation_2017,zhang_physical_2021}. In the referred experiments, the optimal stoichiometry was found to be about 80\% Mo and 20\% Si, as a result of a trade-off between maintaining the stability of an amorphous phase, which requires Si, and maximizing $T_c$, which would favour electron-dense amorphous Mo. The maximum $T_c$ of MoSi in these experiments was 7.9~K~\cite{zhang_physical_2021}, which is significantly higher than in crystalline Mo$_3$Si, which has almost the same stoichiometry, but $T_c$ of only 1.3~K~\cite{tutuncu_electronic_2010}. According to Ref.~\cite{zhang_physical_2021}, $T_c$ of amorphous co-sputtered MoSi decreases rapidly in Mo-poor MoSi when the fraction of Mo is below 45\%. Other methods for fabricating disordered MoSi and other silicides include mixing a deposited metal film into a silicon substrate by ion irradiation~\cite{tsaur_ionbeaminduced_1979} or by annealing, a technique that was used in the observation of QPS~\cite{lehtinen_superconducting_2017}.

The critical temperature of polycrystalline, moderately disordered ($R_\square \sim 10$~$\Omega$ for $\sim 100$~nm film) NbN films films have been found to be relatively insensitive to ion or neutron irradiation~\cite{juang_effects_1988}. However, ion irradiation of superconducting nitrides has recently been utilized for the fabrication of various components or devices, e.g., helium irradiation for superconducting nanowires from NbN~\cite{martinez_superconducting_2020,zhang_saturating_2019} and for highly tuneable Josephson junctions from NbTiN~\cite{ruhtinas_highly_2023}. Argon irradiation has also been used for tuning the local critical current density of NbTiN superconductor wide strip photon detectors~\cite{yabuno_superconducting_2023}.

Recently, it was shown that focused ion beam (FIB) irradiation by gallium or helium ions increases $T_c$, and $R_\square$ of MoSi formed by thermal annealing~\cite{mykkanen_enhancement_2020}. However, the FIB method is a relatively slow direct-writing process, with timescale proportional to the processed area. In the present work, we focus on broad beam ion irradiation treatment that allows wafer-scale processing. We perform a more systematic study of MoSi, including structural imaging, and extend the study into several other materials of interest.

\section{Materials and methods}
\label{sec_materials}

\begin{figure}[h]
    \includegraphics[width=\linewidth]{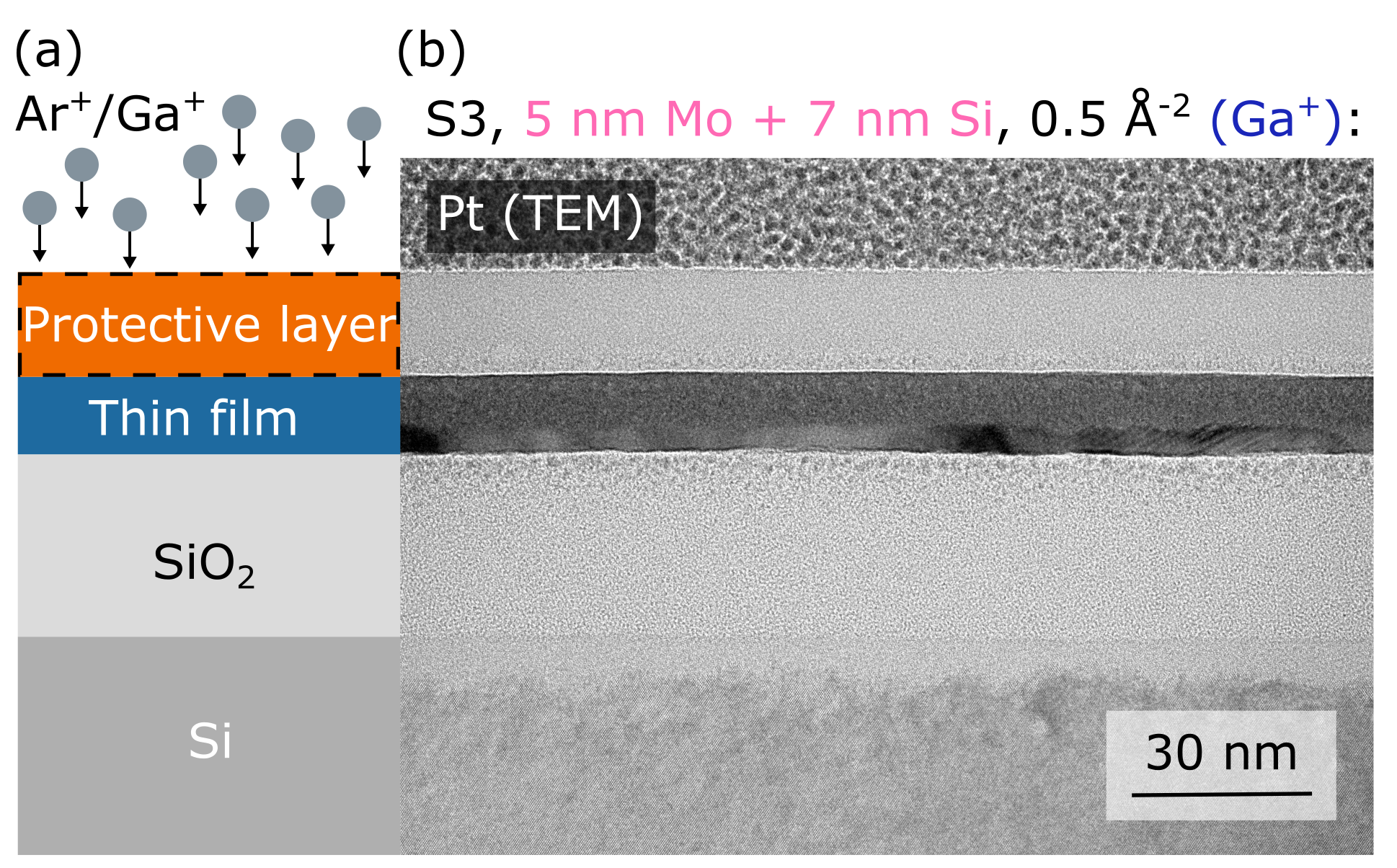}
    \caption{(a) Illustration of the wafer structure and the ion irradiation process. (b) Transmission electron microscope image of MoSi film S3. The layer of platinum on top of the protective layer was deposited for the TEM imaging and is not a part of the actual fabrication process. The title describes the nominal thicknesses of the sputtered Mo and Si layers as well as the ion fluence.}
    \label{wafer_illustration}
\end{figure}

Figure~\ref{wafer_illustration} illustrates our process and wafer structure, and shows the corresponding transmission electron microscope (TEM) image of a (non-ideal) MoSi wafer taken after the ion irradiation treatment. The thin films were sputtered on top of a silicon wafer terminated by a silicon dioxide layer. As the next step we protected some of the wafers with a dielectric layer, because thin films might be sensitive to ageing due to oxidation and wafer-scale processing may require significant shelf life. We experimented two dielectrics: $20~\mathrm{nm}$ SiO$_\mathrm{2}$ grown by plasma enhanced chemical vapor deposition (PECVD) or $15~\mathrm{nm}$ of Al$_\mathrm{2}$O$_\mathrm{3}$ grown by atomic layer deposition (ALD). If necessary, both layers can also be later removed by selective etching.

The wafers were irradiated with either argon or gallium ions. As a noble gas, argon may irradiate without any doping of the superconducting film. On the other hand, the atomic mass of gallium ions yield a closer match for most superconductors, which results in more efficient momentum transfer per impact, and allows smaller ion fluences. Gallium also provided promising results in Ref.~\cite{mykkanen_enhancement_2020} without any evidence of doping problems. Due to the inevitably large number of other variables in our study, we kept the ion acceleration voltage \emph{qualitatively} constant for all wafers by choosing values that maximize the fraction of kinetic energy deposited into the superconduting film. For this purpose, we performed Monte Carlo simulations (not shown) to optimize the voltage for each combination of wafer stack and ion. The resulting different \emph{quantitative} acceleration voltages are reported in the supplementary. Due to this approach, the majority of the irradiation ions are deposited into the film. We report the irradiation fluences in units of \AA$^{-2} = 10^{16}$~cm$^{-2}$ to provide intuitive into the amount of irradiated ions on the atomic scale.

\begin{table}
\caption{Overview of thin film materials, their nominal sputtered thicknesses (those of Mo+Si for MoSi), protective layers, and ions for irradiation.}\label{table_materials} 
\begin{ruledtabular}
\begin{tabular}{c c c c}
Material        & Thickness (nm)    & Protective layer  & Ion \\ \hline
\multicolumn{4}{l}{Single-elements} \\ \hline
Al              & $14$      & none                              & $\mathrm{Ar^+}$  \\ 
Nb              & $20$      & Al$_\mathrm{2}$O$_\mathrm{3}$/none             & $\mathrm{Ar^+}$  \\ \hline
\multicolumn{4}{l}{Nitrides} \\ \hline
TiN             & $10$      & Al$_\mathrm{2}$O$_\mathrm{3}$/none             & $\mathrm{Ar^+}$  \\
NbN             & $10$      & Al$_\mathrm{2}$O$_\mathrm{3}$/none             & $\mathrm{Ar^+}$  \\ \hline
\multicolumn{4}{l}{Silicides} \\ \hline
MoSi (M-series) & $10+7$     & SiO$_\mathrm{2}$/Al$_\mathrm{2}$O$_\mathrm{3}$& $\mathrm{Ar^+}$/$\mathrm{Ga^+}$ \\
MoSi (S-series) & $5+7$     & Al$_\mathrm{2}$O$_\mathrm{3}$                  & $\mathrm{Ar^+}$/$\mathrm{Ga^+}$ \\
MoSi (S-series) & $3+7$     & Al$_\mathrm{2}$O$_\mathrm{3}$                  & $\mathrm{Ar^+}$/$\mathrm{Ga^+}$  
\end{tabular}
\end{ruledtabular}
\end{table}

Table~\ref{table_materials} provides an overview of the thin film materials, their targeted sputtering thicknesses, protective layers, and the ions of the irradiation treatment. For MoSi, we performed a study of the effect of ion fluence for Mo-rich stoichiometry (M-series wafers), and a smaller number of process variants for Si-rich stoichiometry (S-series). 

Nitrides were deposited by reactive sputtering of Nb or Ti in a flow of argon and nitrogen. For TiN, we used a recipe that produces $T_c\approx 3~\mathrm{K}$. For NbN, the maximum $T_c$ of $14~\mathrm{K}$ is obtained with 1:1 stoichiometry~\cite{linzen_structural_2017}, but in this work, we aimed at high $R_\square$ by utilizing two higher nitrogen flow values. Below we denote these materials as NbN and NbN$^*$, and their expected values of $T_c$ are 7~K and 6~K, respectively.

For MoSi films, first a layer of molybdenum and then a layer of silicon were deposited, which allowed tuning the targeted stoichiometry by changing the ratio of the film thicknesses. The thickness estimates are based on longer depositions using the same sputtering parameters, and measuring the resulting thicker films. However, the short deposition times of thin films increase irreproducibility of their thicknesses. We estimate the stoichiometries of about 65\% Mo + 35\% Si, 48\% Mo + 52\% Si, and 35\% Mo + 65\% Si for our film variants with $10~\mathrm{nm}$, $5~\mathrm{nm}$, and $3~\mathrm{nm}$ of Mo (and $7~\mathrm{nm}$ of Si), respectively. The MoSi compounds were formed by annealing in nitrogen environment at $600~^\circ\mathrm{C}$ for $15$ minutes after the deposition of the protective layer. The same temperature (but only for 10 min) was used in Ref.~\cite{mykkanen_enhancement_2020} and was expected to yield a compromise between allowing Mo to diffuse into amorphous Si, but not allowing Si to crystallize before that, which would prevent forming the compound~\cite{liang_interfacial_1996}. However, the optimal annealing of MoSi may depend on many details of the fabrication process, including the thicknesses of the initial Mo and Si layers, and such optimization is beyond the scope of the present work.

In total, we produced 32 irradiated wafers and varied materials, protective layers, and ions and their fluence. We also fabricated reference wafers with the same processes excluding ion irradiation for Al, Nb, TiN, and NbN. We measured $R_\square$ of all wafers at room temperature and imaged most of them with a scanning electron microscope (SEM). We used data from this fast characterization to select an illustrative set of samples for the more elaborate cryogenic characterization (presented in Sec.~\ref{Rs_Tc_section}). Based on all these data, a set of 7 MoSi films were selected for TEM imaging (Sec.~\ref{Sec_imaging}): identical MoSi films with varied ion fluence, films that seemed visually defective in SEM, MoSi with highest $T_c$, and MoSi films that were promising in all of the initial measurements.

\section{Material mixing\label{Sec_imaging}}

\begin{figure*}
    \includegraphics[width=\linewidth]{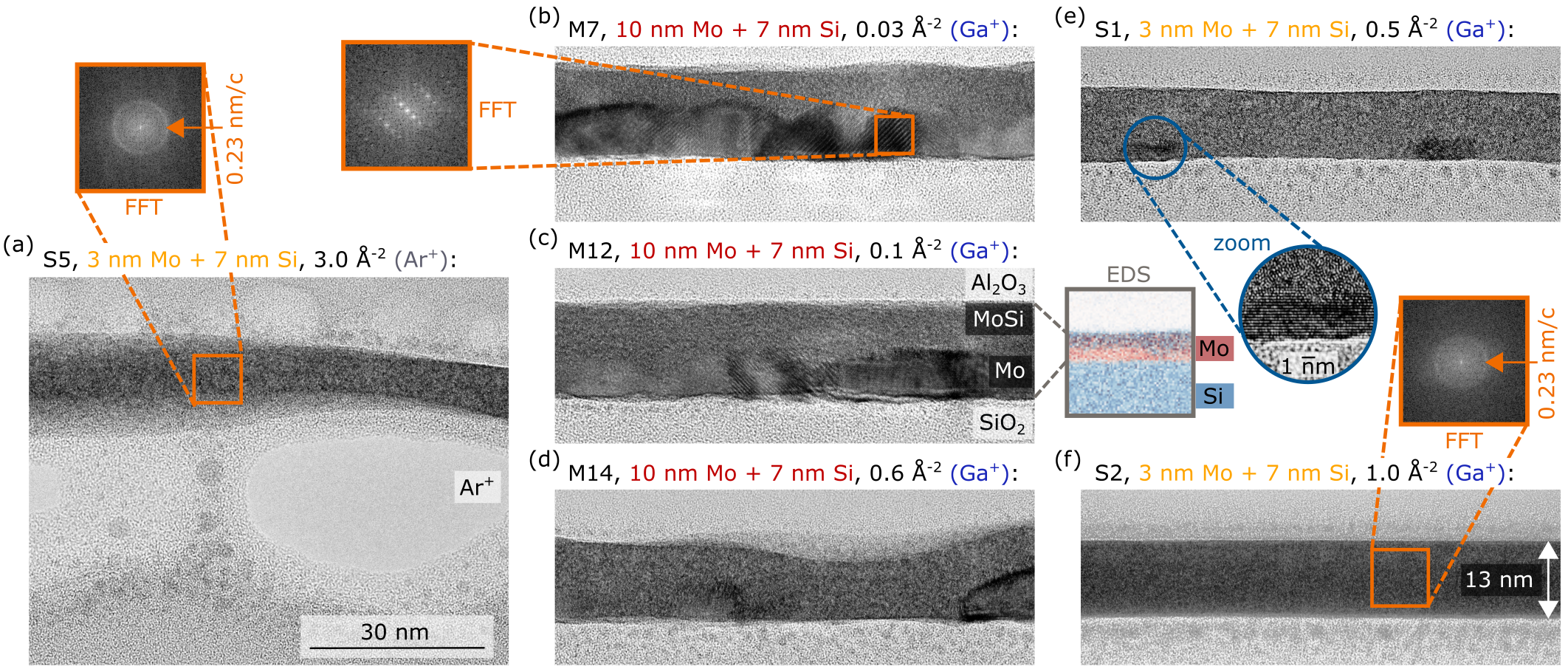}
    \caption{Transmission electron microscope images of MoSi films S5, M7, M12, M14, S1, and S2 (a--f, respectively). Images (a--f) are on the same scale shown in (a) and (f). The title of each image describes the nominal thicknesses of the sputtered Mo and Si layers, and the fluence and ion of the irradiation treatment. The layer stack is the same in each figure: the bright bottom layer is SiO$_\mathrm{2}$ (oxide capping of the silicon wafer) and the bright top layer is of Al$_\mathrm{2}$O$_\mathrm{3}$ deposited to protect the darker layer consisting of Mo and Si. The insets in (a--b) and (f) show the FFT analysis from the region shown with the orange square. The inset of (c) shows the EDS data of Mo and Si (blue for Si, red for Mo) combined. The grey dashed lines illustrate where the MoSi layer is shown in the EDS inset. The inset in (e) shows the zoom of the region shown with the blue circle. \label{TEM_MoSi}}
\end{figure*}

Figure~\ref{TEM_MoSi}(a) shows the TEM image of a Mo-poor MoSi film irradiated with argon, which yielded the highest $T_c=5.7$~K of all our MoSi films (see Sec.~\ref{Rs_Tc_section} below). There are gas pockets both in the substrate and protective layer, which deform the MoSi film and cause variation of its thickness. Since the SEM images (not shown) of all argon irradiated MoSi wafers indicated visually uneven surface, we expect that also they had gas pockets. The energy-dispersive X-ray spectroscopy (EDS) data (not shown) relate the gas pockets to argon. The film is continuous, with grains not resolvable with TEM. Fast Fourier transform (FFT) analysis shows a broad ring at 0.23~nm/cycle. This suggests that the film is amorphous with short-range order corresponding to the nearest neighbors distance.

Figures~\ref{TEM_MoSi}(b--d) show TEM images of Mo-rich MoSi films treated with an increasing fluence of $\mathrm{Ga^+}$ ions. The films are heterogeneous and have a bilayer structure, with a polycrystalline bottom and an amorphous top layer. Inset of FFT analysis of the bottom layer confirms the crystallinity of the grains. Inset of EDS data indicates that the bottom layer consists mostly of molybdenum whereas the top layer is a MoSi compound. Higher fluences of gallium reduce the amount of crystalline Mo.  Figure~\ref{wafer_illustration}(b) shows the TEM image of Mo-poor film (S3) with $5~\mathrm{nm}$ of Mo. The image shows a bilayer structure, similarly as for the Mo-rich films of Figs.~\ref{TEM_MoSi}(b--c).

Figures~\ref{TEM_MoSi}(e--f) show TEM images of Mo-poor MoSi films (S1--S2) with $3~\mathrm{nm}$ of Mo. The thinner layer of molybdenum yields a better intermixing with the silicon top layer, compare Figs.~\ref{wafer_illustration}(b) and \ref{TEM_MoSi}(e) with the same gallium fluence of 0.5~\AA$^{-2}$. Increasing the fluence to 1.0~\AA$^{-2}$ leads to a continuous, homogeneous film of molybdenum and silicon, without TEM-resolvable grains. The FFT analysis shows a broad ring at 0.23~nm/cycle, suggesting that the film is amorphous with short-range order.

Figures~\ref{wafer_illustration} and \ref{TEM_MoSi} indicate that our annealing process did not result in homogeneous mixing of Mo and Si especially for the Mo-rich stoichiometries, but that ion irradiation improves it. Ion irradiation may thus allow the fabrication of homogeneous, amorphous MoSi without co-sputtering technology~\cite{bosworth_amorphous_2015,banerjee_characterisation_2017,zhang_physical_2021}. We expect that for a two-layer structure, the mixing improves if the atomic mass of the irradiation ion is closer to that of the top layer rather than to the bottom one. In this respect Ar$^+$ (40 u) is more optimal than Ga$^+$ (70 u) in our case where Si (28 u) is on top of Mo (96 u). Finally, we note that we did not observe significant gas pockets in TEM images of NbN films (not shown). The present amount of data thus does not rule out the use of $\mathrm{Ar^+}$ ions for irradiation, but it would also be interesting to explore noble gases with higher atomic masses.

An upper limit for the gallium doping of our MoSi films can be obtained by assuming that all ions remain in the superconducting film. For the thinnest film and largest gallium fluence (3~nm Mo + 7~nm Si, 1~\AA$^{-2}$), this limit yields the stoichiometry of 30\% Mo, 54\% Si and 16\% Ga (35\% Mo, 65\% Si before irradiation).

\section{Sheet resistance $R_{\square}$ and critical temperature $T_c$\label{Rs_Tc_section}}

We measured the room temperature sheet resistance $R_{\square,\mathrm{RT}}$ with a wafer prober. For cryogenic measurements, we cleaved the wafers into small chips that were fully covered with the film on the top surface. The relative temperature dependence of $R_\square$
down to the base temperature of 0.3~K of our cryostat is recorded through measurements with 4-probe geometry realized by wire bonding. Combining these data sets yields sheet resistance as a function of temperature, $R_{\square}(T)$.

\begin{figure}
    \includegraphics[width=\linewidth]{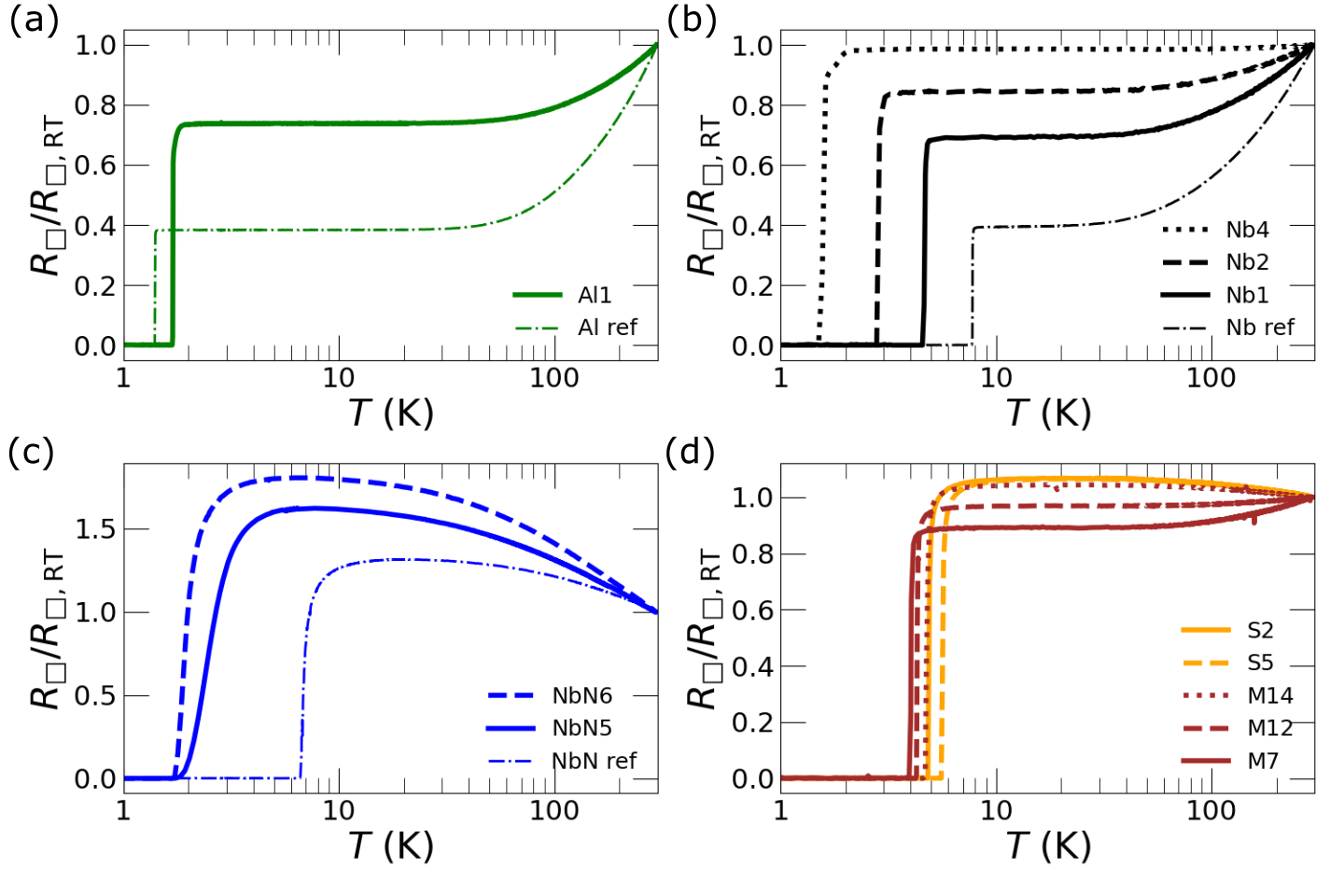}
    \caption{Normalized sheet resistance $R_{\square}/R_{\square,\mathrm{RT}}$ as a function of temperature for (a) Al, (b) Nb, (c) NbN, and (d)  S- and M-series MoSi films. In (a--c), "ref" labels reference wafers that were not irradiated.}
    \label{Tc_sweep_figure}
\end{figure}

\begin{figure*}[h!t]
    \includegraphics[width=\linewidth]{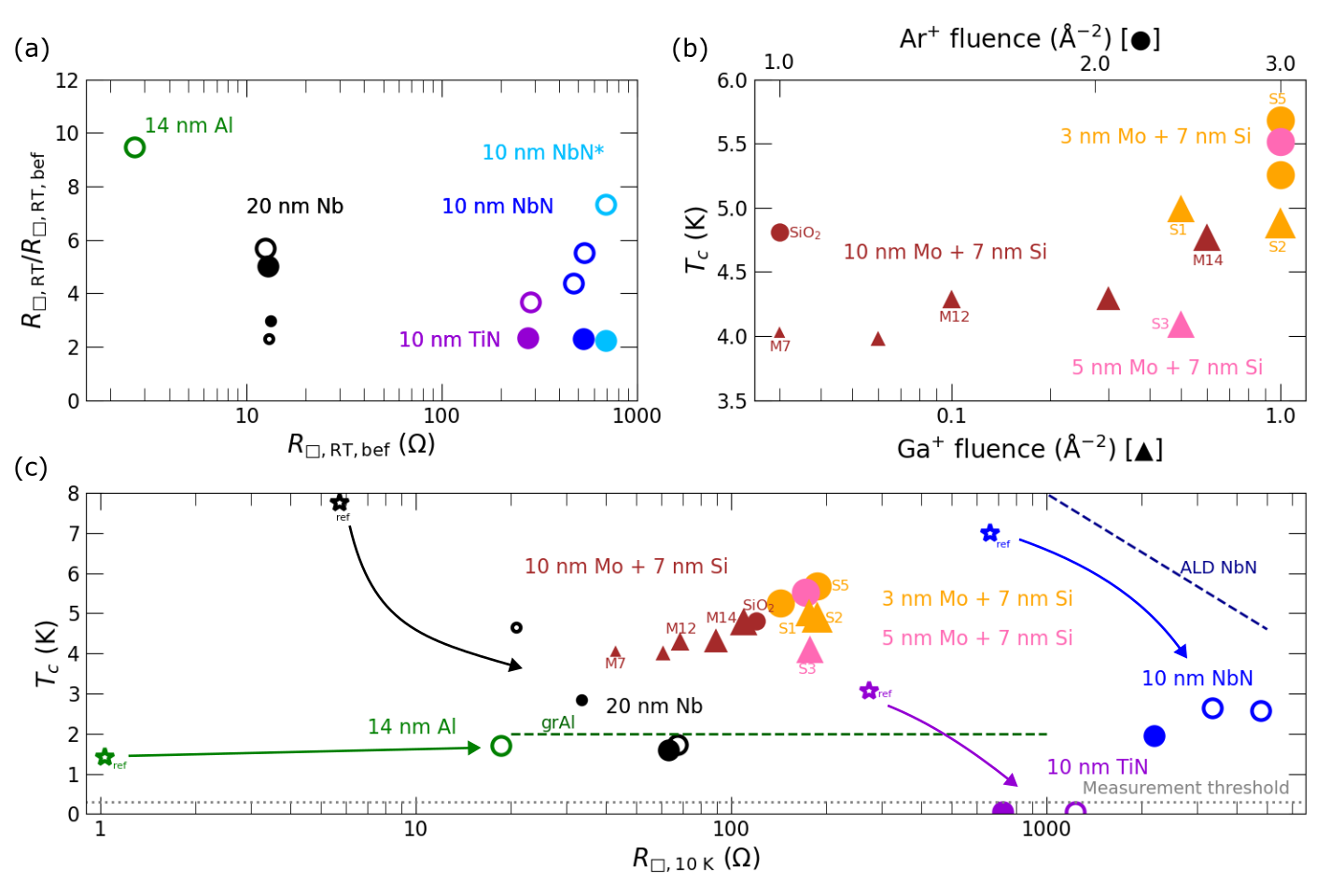}
    \caption{The relationships between ion fluence, sheet resistance, and superconducting transition temperature measured down to measurement threshold of 0.3~K. The material labels are the nominal thicknesses of the sputtered films and different colors are used for different materials. The size of the markers illustrates the magnitude of fluence (for argon and gallium independently). The films treated with gallium and argon are shown with triangle and circle markers, respectively. Films that were TEM imaged are labeled. Film that had a SiO$_\mathrm{2}$ protective layer is labeled (other films with a protective layer had Al$_\mathrm{2}$O$_\mathrm{3}$). The empty markers mean that those wafers did not have a protective layer. (a) Increase in sheet resistance due to ion irradiation treatment ($R_{\square, \mathrm{RT}}/R_{\square,\mathrm{RT,bef}}$) as a function of $R_{\square,\mathrm{RT,bef}}$, which is the sheet resistance for otherwise finished wafer but without the ion irradiation treatment. The lighter blue NbN, also marked with asterisk, presents the NbN wafers with higher nitrogen flow (NbN7-8). (b) $T_c$ as a function of argon or gallium ion fluence for M- and S-series of MoSi. (c) $T_c$ as a function of $R_{\square, 10~\mathrm{K}}$. Star markers show (in some cases average) values for wafers which did not undergo ion irradiation treatment, also labeled as "ref". The arrows illustrate the change in parameters due to ion irradiation. Examples of parameter ranges with alternative techniques, granular aluminium (grAl)~\cite{levy-bertrand_electrodynamics_2019} and ALD-grown NbN~\cite{linzen_structural_2017}, are shown with dashed green and blue lines, respectively.}
    \label{Rs_Tc_dose_figure}
\end{figure*}

In this section, Fig.~\ref{Tc_sweep_figure} shows the most interesting data on $R_{\square}(T)$ for Al, Nb, NbN, and MoSi. A more comprehensive set of data based on the analysis of all $R_{\square}(T)$ measurements is collated into table~\ref{TEM_MoSi_NbN_table} (see the supplementary for complete data). Table~\ref{TEM_MoSi_NbN_table} also compares irradiated films to values obtained from reference wafers without irradiation. We conclude this section by illustrating the dependencies of table~\ref{TEM_MoSi_NbN_table} in Fig.~\ref{Rs_Tc_dose_figure}.

\begin{table*}
\caption{Electrical measurement data for films presented on Fig.~\ref{Rs_Tc_dose_figure}. The columns describe wafer labels, sputtering deposition thicknesses (Mo+Si for MoSi), protective layers, ions and fluences of the ion irradiation treatments, room temperature sheet resistances before and after ion irradiation treatment, corresponding increases in room temperature sheet resistances, sheet resistances at $10~\mathrm{K}$, critical temperatures, and the changes in critical temperatures compared to the reference values measured from the wafers without ion irradiation. Hyphen means that the corresponding measurement was not performed. The MoSi films that are shown in Figs.~\ref{wafer_illustration}, ~\ref{TEM_MoSi} are indicated in bold.
\label{TEM_MoSi_NbN_table}}
\begin{ruledtabular}
\begin{tabular}{l c c c c c c c c c c}
Label           & Dep.~thick.~(nm)                  & Prot.                          & Ion             & Fluence (\AA$^{-2}$) & $R_{\square, \mathrm{RT, bef}}$ $(\Omega)$ & $R_{\square, \mathrm{RT}}$ $(\Omega)$ & $R_{\square, \mathrm{RT}}$/$R_{\square, \mathrm{RT, bef}}$ & $R_{\square, \mathrm{10~K}}$ $(\Omega)$ & $T_c$ (K) & $T_c/T_{c,\mathrm{ref}}$ \\ \hline
Al1             & 14      & none                           & $\mathrm{Ar^+}$ & 1.5     & 2.7   & 25    & 9.5   & 19    & 1.7 & 1.2  \\ \hline
Nb1             & 20      & none                           & $\mathrm{Ar^+}$ & 0.5     & 13    & 30    & 2.3   & 21    & 4.7 & 0.60 \\
Nb2             & 20      & Al$_\mathrm{2}$O$_\mathrm{3}$  & $\mathrm{Ar^+}$ & 0.5     & 13    & 40    & 3.0   & 33    & 2.8 & 0.37 \\
Nb3             & 20      & none                           & $\mathrm{Ar^+}$ & 1.5     & 13    & 71    & 5.7   & 68    & 1.7 & 0.22 \\
Nb4             & 20      & Al$_\mathrm{2}$O$_\mathrm{3}$  & $\mathrm{Ar^+}$ & 1.5     & 13    & 64    & 5.0   & 63    & 1.6 & 0.20 \\ \hline
NbN3            & 10      & none                           & $\mathrm{Ar^+}$ & 1.5     & 470   & 2100  & 4.4   & 3300  & 2.6 & 0.38 \\ 
NbN5            & 10      & none                           & $\mathrm{Ar^+}$ & 1.5     & 530   & 3000  & 5.5   & 4800  & 2.6 & 0.37 \\
NbN6            & 10      & Al$_\mathrm{2}$O$_\mathrm{3}$  & $\mathrm{Ar^+}$ & 1.5     & 530   & 1200  & 2.3   & 2200  & 2.0 & 0.28 \\
NbN7$^*$        & 10      & none                           & $\mathrm{Ar^+}$ & 1.5     & 690   & 5100  & 7.3   & --    & --  & --   \\ 
NbN8$^*$        & 10      & Al$_\mathrm{2}$O$_\mathrm{3}$  & $\mathrm{Ar^+}$ & 1.5     & 690   & 1500  & 2.2   & --    & --  & --   \\ \hline
TiN1            & 10      & none                           & $\mathrm{Ar^+}$ & 1.5     & 290   & 1000  & 3.7   & 1200  & \textless 0.3 & \textless 0.1 \\
TiN2            & 10      & Al$_\mathrm{2}$O$_\mathrm{3}$  & $\mathrm{Ar^+}$ & 1.5     & 280   & 640   & 2.3   & 720   & \textless 0.3 & \textless 0.1 \\  \hline
M1              & $10+7$  & SiO$_\mathrm{2}$               & $\mathrm{Ar^+}$ & 1.0     & --    & 120   & --    & 120   & 4.8 & --   \\
\textbf{M7}     & $10+7$  & Al$_\mathrm{2}$O$_\mathrm{3}$  & $\mathrm{Ga^+}$ & 0.03    & --    & 48    & --    & 43    & 4.0 & --   \\
M11             & $10+7$  & Al$_\mathrm{2}$O$_\mathrm{3}$  & $\mathrm{Ga^+}$ & 0.06    & --    & 65    & --    & 61    & 4.0 & --   \\
\textbf{M12}    & $10+7$  & Al$_\mathrm{2}$O$_\mathrm{3}$  & $\mathrm{Ga^+}$ & 0.1     & --    & 71    & --    & 69    & 4.3 & --   \\
M13             & $10+7$  & Al$_\mathrm{2}$O$_\mathrm{3}$  & $\mathrm{Ga^+}$ & 0.3     & --    & 88    & --    & 90    & 4.3 & --   \\
\textbf{M14}    & $10+7$  & Al$_\mathrm{2}$O$_\mathrm{3}$  & $\mathrm{Ga^+}$ & 0.6     & --    & 110   & --    & 110   & 4.8 & --   \\
\textbf{S1}     & $3+7$   & Al$_\mathrm{2}$O$_\mathrm{3}$  & $\mathrm{Ga^+}$ & 0.5     & --    & 170   & --    & 180   & 5.0 & --   \\ 
\textbf{S2}     & $3+7$   & Al$_\mathrm{2}$O$_\mathrm{3}$  & $\mathrm{Ga^+}$ & 1.0     & --    & 180   & --    & 190   & 4.9 & --   \\ 
\textbf{S3}     & $5+7$   & Al$_\mathrm{2}$O$_\mathrm{3}$  & $\mathrm{Ga^+}$ & 0.5     & --    & 170   & --    & 180   & 4.1 & --   \\
S4              & $3+7$   & Al$_\mathrm{2}$O$_\mathrm{3}$  & $\mathrm{Ar^+}$ & 3.0     & --    & 140   & --    & 140   & 5.3 & --   \\
\textbf{S5}     & $3+7$   & Al$_\mathrm{2}$O$_\mathrm{3}$  & $\mathrm{Ar^+}$ & 3.0     & --    & 180   & --    & 190   & 5.7 & --   \\ 
S6              & $5+7$   & Al$_\mathrm{2}$O$_\mathrm{3}$  & $\mathrm{Ar^+}$ & 3.0     & --    & 160   & --    & 170   & 5.5 & --   \\
\end{tabular}
\end{ruledtabular}
\end{table*}

The general trend of Fig.~\ref{Tc_sweep_figure} is that the residual resistivity ratio (RRR) $R_{\square,\mathrm{RT}}/R_{\square,\mathrm{LT}}$ between room-temperature (RT) and normal-state low-temperature (LT) resistance is higher for the less disordered films (compare to table~\ref{TEM_MoSi_NbN_table} for the resistance values). Panels (a--c) show that ion irradiation increases $T_c$ of Al, but decreases those of Nb and NbN. The transitions of NbN, the most disordered of our films, are significantly more smeared than for the others. Figure~\ref{Tc_sweep_figure}(d) shows that RRR$>1$ for Mo-rich films that contain a layer of polycrystalline Mo, and RRR$<1$ for the better mixed S series films, see Fig.~\ref{TEM_MoSi}. Improving material mixing also increases $T_c$. Comparison to non-irradiated MoSi wafers is not relevant due to the imperfect mixing of Mo and Si in such films.

Table~\ref{TEM_MoSi_NbN_table} shows that ion irradiation and increasing ion fluence increases $R_\square$ for all materials. For NbN, there is a significant difference in $R_\square$ also between films with and without a protective layer. These results are also illustrated in Fig.~\ref{Rs_Tc_dose_figure}(a) for Al, Nb, NbN and TiN.

For MoSi, interpretation of the results is less straightforward, since most of the films are not homogeneous. The increase of ion fluence reduces the amount of polycrystalline Mo (see Figs.~\ref{wafer_illustration}--\ref{TEM_MoSi}) which we expect to dominate conductivity if there is a continuous layer. In contrast, $T_c$ measurement of a multilayer stack yields the highest $T_c$ of the stack, but the other layers can still affect the result through the inverse proximity effect. Figure~\ref{Rs_Tc_dose_figure}(b) illustrates $T_c$ of M- and S-series MoSi as a function of ion fluence. The values of $T_c$ are between $4~\mathrm{K}$ and $6~\mathrm{K}$, and the highest values are obtained for the most disordered Mo-poor films. This result is in ostensible disagreement with
Ref.~\cite{zhang_physical_2021}, where highest values of $T_c$ were obtained with Mo-rich films, but our results on Mo-rich films were likely affected by the imperfect mixing. However, our highest $T_c$ for MoSi, 5.7~K (film S5, irradiated with argon) is remarkably high compared to the value 2.8~K obtained for the same stoichiometry in Ref.~\cite{zhang_physical_2021} for co-sputtered, presumably amorphous MoSi. 

Figure~\ref{Rs_Tc_dose_figure}(b) shows one film that had SiO$_\mathrm{2}$ as a protective dielectric (see Table~\ref{TEM_MoSi_NbN_table}, wafer M1). All other films in Fig.~\ref{Rs_Tc_dose_figure}(b) were protected with Al$_\mathrm{2}$O$_\mathrm{3}$. However, several films similar to M1, protected with SiO$_\mathrm{2}$ and irradiated with the same fluence had reproducible values of
$R_{\square, \mathrm{RT}}$ between 98--120$~\mathrm{\Omega}$, see the supplementary. The $T_c$ of wafer M1 is also the same as the highest $T_c$ measured from gallium-treated M-series films with Al$_\mathrm{2}$O$_\mathrm{3}$ layer. Hence we notice no difference between the two protective layer materials.

Finally, Fig.~\ref{Rs_Tc_dose_figure}(c) shows the map between $T_c$ and $R_{\square, 10~\mathrm{K}}$ of MoSi, NbN, TiN, Nb, and Al wafers. Reference values of non-irradiated wafers are included for Al, Nb, TiN, and NbN. The increase of $T_c$ in Al and the corresponding decrease in Nb as a function of increased disorder are in qualitative agreement with experiments on granular Al~\cite{levy-bertrand_electrodynamics_2019} and Nb~\cite{bose_mechanism_2005}, but obtained with an entirely different fabrication method. A significantly smaller decrease of $T_c$, 0.16~K, has been observed in Nb films irradiated with protons~\cite{tanatar_anisotropic_2022}, which was in agreement with theoretical calculations~\cite{zarea_effects_2023}. However, the proton-irradiated Nb films had an order of magnitude smaller resistivity than films Nb3--4. Strong decrease of $T_c$ as a function of ion fluence and resistivity makes this elemental material less promising as a disordered superconductor. On the other hand, lithography-defined irradiation may allow Nb structures of different $T_c$ and $L_\square$ in the same layer.

Ion irradiation yields a significant decrease of $T_c$ for both nitrides. In irradiated TiN, we did not observe superconductivity at and above $0.3~\mathrm{K}$. The effect of ion irradiation on $T_c$ of our 10~nm NbN films is significantly stronger ($\sim 70$\%) than in thicker and less disordered films studied in Ref.~\cite{juang_effects_1988} ($\sim 10$\%).

We studied Nb, NbN, and TiN films both with and without protective layers. Our experiments do not indicate any noticeable effect of the protective layer on the measured $T_c$, but the increase of $R_\square$ was significantly higher for non-protected nitrides. However, this effect may also be attributed to oxidation, i.e., the limited shelf-life of thin nitride films.

We consider Fig.~\ref{Rs_Tc_dose_figure}(c) also as an application-oriented map into disordered superconductors, from which one could pick suitable materials for each purpose. Since ion irradiation is in principle suitable for increasing disorder of any superconducting thin film, future research on other materials may help to fill the empty parts of the map. As examples of alternative techniques, tuning the oxidation of granular aluminium provides a wide range of $R_\square$ at $T_c\approx 2$~K~\cite{levy-bertrand_electrodynamics_2019}, while reducing the thickness of ALD-grown NbN can provide values from about $T_c\approx 14$~K and $R_\square \approx 60$~$\Omega$ to $T_c\approx 5$~K and $R_\square \approx 5$~k$\Omega$~\cite{linzen_structural_2017}. These value ranges of the alternative techniques are illustrated in Fig.~\ref{Rs_Tc_dose_figure}(c) by the green and blue lines, respectively.

\section{Discussion and conclusions}

We present a wafer-scale method for increasing disorder in superconducting thin films by using broad beam ion irradiation. Our treatment can be applied independently of the material and its fabrication method, but a permanent effect requires a disordered phase that is stable at room temperature. In contrast to most other techniques, ion irradiation is also expected to allow limiting the treatment to certain parts of the thin film by lithography, which enables pattern transfer of disorder-induced superconducting properties within the same metallization layer.

We used both gallium and argon ions and studied various single-element (Al, Nb) and compound materials (NbN, TiN, MoSi). Both ions increased sheet resistance of the films, but in some cases, argon yielded gas pockets that can be detrimental for applications. As a result of ion irradiation, the critical temperature increased in MoSi and Al, but decreased in TiN, NbN, and Nb.

Both ions were used to produce a MoSi film that is amorphous with some short-range order. We obtained significantly higher $T_c$ (5.7~K) for Mo-poor MoSi film (35\% Mo, 65\% Si) than in literature (2.8~K)~\cite{zhang_physical_2021}. Our Mo-rich MoSi films had the tendency of forming a two-layer structure of polycrystalline Mo and amorphous MoSi. Further research on the effect of stoichiometry would benefit from better mixing of the material before ion irradiation, which could be achieved, e.g., through co-sputtering from Mo and Si targets or sputtering from a MoSi compound target.

The ion irradiation method enables tuning of $R_{\square}$ and $T_c$ of superconducting thin films, potentially expanding the range of promising materials for different devices in quantum technology. In this work, we have demonstrated amorphous and uniform MoSi that is promising for superconducting nanowire single-photon detectors as well as tuning of parameters for Al, Nb, and NbN thin films.

\section*{Supplementary material}

A table that shows information about all wafers of this work is provided in the supplementary material. The first columns contain the wafer labels, sputtered materials, protective layers, and sheet resistances in room temperature after deposition. The next columns describe the ion irradiation treatment parameters, i.e., ions, fluences, and acceleration voltages. The last columns show the sheet resistances in room temperature after ion irradiation treatment, sheet resistances in 10~K, and critical temperatures. The $T_c$ value is defined as the temperature where $R(T) = 0.5\times R_{10 \mathrm{K}}$. Sheet resistances in room temperature have been measured by a four point probe in multiple positions of the wafer and the value in the table is the average of those measurements, shown with two significant figures. The asterisk in some NbN wafer labels correspond to different sputtering parameters, see Sec.~\ref{sec_materials}.

\begin{acknowledgments}
We thank J. A. Sauls for useful discussions. For funding of our research project, we acknowledge the European Union’s Horizon 2020 Research and Innovation Programme under the Grant Agreement Nos.~862660/Quantum e-leaps, 899558/aCryComm, 766853/EFINED, and ECSEL programme 101007322/MatQu. This project has also received funding from Business Finland through Quantum Technologies Industrial (QuTI) project No.~128291 and from Research Council of Finland through Grant Nos.~310909 and 350220 and Finnish Quantum Flagship project 359284. This work was performed as part of the Research Council of Finland Centre of Excellence program (projects 336817, 336819, 352934, and 352935).  We also acknowledge funding from an internal strategic innovation project of VTT related to the development of quantum computing technologies. This research was supported by the Scientific Service Units of IST Austria through resources provided by Electron Microscopy Facility. J. Senior acknowledges funding from the European Union’s Horizon 2020 Research and Innovation Programme under the Marie Sk\l odowska-Curie Grant Agreement No. 754411. A. Ronzani acknowledges funding from Research Council of Finland (Research Fellowship project 356542).
\end{acknowledgments}


\begin{thebibliography}{43}%
\makeatletter
\providecommand \@ifxundefined [1]{%
 \@ifx{#1\undefined}
}%
\providecommand \@ifnum [1]{%
 \ifnum #1\expandafter \@firstoftwo
 \else \expandafter \@secondoftwo
 \fi
}%
\providecommand \@ifx [1]{%
 \ifx #1\expandafter \@firstoftwo
 \else \expandafter \@secondoftwo
 \fi
}%
\providecommand \natexlab [1]{#1}%
\providecommand \enquote  [1]{``#1''}%
\providecommand \bibnamefont  [1]{#1}%
\providecommand \bibfnamefont [1]{#1}%
\providecommand \citenamefont [1]{#1}%
\providecommand \href@noop [0]{\@secondoftwo}%
\providecommand \href [0]{\begingroup \@sanitize@url \@href}%
\providecommand \@href[1]{\@@startlink{#1}\@@href}%
\providecommand \@@href[1]{\endgroup#1\@@endlink}%
\providecommand \@sanitize@url [0]{\catcode `\\12\catcode `\$12\catcode
  `\&12\catcode `\#12\catcode `\^12\catcode `\_12\catcode `\%12\relax}%
\providecommand \@@startlink[1]{}%
\providecommand \@@endlink[0]{}%
\providecommand \url  [0]{\begingroup\@sanitize@url \@url }%
\providecommand \@url [1]{\endgroup\@href {#1}{\urlprefix }}%
\providecommand \urlprefix  [0]{URL }%
\providecommand \Eprint [0]{\href }%
\providecommand \doibase [0]{http://dx.doi.org/}%
\providecommand \selectlanguage [0]{\@gobble}%
\providecommand \bibinfo  [0]{\@secondoftwo}%
\providecommand \bibfield  [0]{\@secondoftwo}%
\providecommand \translation [1]{[#1]}%
\providecommand \BibitemOpen [0]{}%
\providecommand \bibitemStop [0]{}%
\providecommand \bibitemNoStop [0]{.\EOS\space}%
\providecommand \EOS [0]{\spacefactor3000\relax}%
\providecommand \BibitemShut  [1]{\csname bibitem#1\endcsname}%
\let\auto@bib@innerbib\@empty
\bibitem [{\citenamefont {Anderson}(1959)}]{anderson_theory_1959}%
  \BibitemOpen
  \bibfield  {author} {\bibinfo {author} {\bibfnamefont {P.~W.}\ \bibnamefont
  {Anderson}},\ }\bibfield  {title} {\enquote {\bibinfo {title} {Theory of
  dirty superconductors},}\ }\href {\doibase 10.1016/0022-3697(59)90036-8}
  {\bibfield  {journal} {\bibinfo  {journal} {Journal of Physics and Chemistry
  of Solids}\ }\textbf {\bibinfo {volume} {11}},\ \bibinfo {pages} {26--30}
  (\bibinfo {year} {1959})}\BibitemShut {NoStop}%
\bibitem [{\citenamefont {Astafiev}\ \emph {et~al.}(2012)\citenamefont
  {Astafiev}, \citenamefont {Ioffe}, \citenamefont {Kafanov}, \citenamefont
  {Pashkin}, \citenamefont {Arutyunov}, \citenamefont {Shahar}, \citenamefont
  {Cohen},\ and\ \citenamefont {Tsai}}]{astafiev_coherent_2012}%
  \BibitemOpen
  \bibfield  {author} {\bibinfo {author} {\bibfnamefont {O.~V.}\ \bibnamefont
  {Astafiev}}, \bibinfo {author} {\bibfnamefont {L.~B.}\ \bibnamefont {Ioffe}},
  \bibinfo {author} {\bibfnamefont {S.}~\bibnamefont {Kafanov}}, \bibinfo
  {author} {\bibfnamefont {Y.~A.}\ \bibnamefont {Pashkin}}, \bibinfo {author}
  {\bibfnamefont {K.~Y.}\ \bibnamefont {Arutyunov}}, \bibinfo {author}
  {\bibfnamefont {D.}~\bibnamefont {Shahar}}, \bibinfo {author} {\bibfnamefont
  {O.}~\bibnamefont {Cohen}}, \ and\ \bibinfo {author} {\bibfnamefont {J.~S.}\
  \bibnamefont {Tsai}},\ }\bibfield  {title} {{\enquote
  {\bibinfo {title} {Coherent quantum phase slip},}\ }}\href {\doibase
  10.1038/nature10930} {\bibfield  {journal} {\bibinfo  {journal} {Nature}\
  }\textbf {\bibinfo {volume} {484}},\ \bibinfo {pages} {355--358} (\bibinfo
  {year} {2012})},\ \bibinfo {note} {number: 7394 Publisher: Nature Publishing
  Group}\BibitemShut {NoStop}%
\bibitem [{\citenamefont {Weitzel}\ \emph {et~al.}(2023)\citenamefont
  {Weitzel}, \citenamefont {Pfaffinger}, \citenamefont {Maccari}, \citenamefont
  {Kronfeldner}, \citenamefont {Huber}, \citenamefont {Fuchs}, \citenamefont
  {Mallord}, \citenamefont {Linzen}, \citenamefont {Il{'}ichev}, \citenamefont
  {Paradiso},\ and\ \citenamefont {Strunk}}]{weitzel_sharpness_2023}%
  \BibitemOpen
  \bibfield  {author} {\bibinfo {author} {\bibfnamefont {A.}~\bibnamefont
  {Weitzel}}, \bibinfo {author} {\bibfnamefont {L.}~\bibnamefont {Pfaffinger}},
  \bibinfo {author} {\bibfnamefont {I.}~\bibnamefont {Maccari}}, \bibinfo
  {author} {\bibfnamefont {K.}~\bibnamefont {Kronfeldner}}, \bibinfo {author}
  {\bibfnamefont {T.}~\bibnamefont {Huber}}, \bibinfo {author} {\bibfnamefont
  {L.}~\bibnamefont {Fuchs}}, \bibinfo {author} {\bibfnamefont
  {J.}~\bibnamefont {Mallord}}, \bibinfo {author} {\bibfnamefont
  {S.}~\bibnamefont {Linzen}}, \bibinfo {author} {\bibfnamefont
  {E.}~\bibnamefont {Il{'}ichev}}, \bibinfo {author} {\bibfnamefont
  {N.}~\bibnamefont {Paradiso}}, \ and\ \bibinfo {author} {\bibfnamefont
  {C.}~\bibnamefont {Strunk}},\ }\bibfield  {title} {\enquote {\bibinfo {title}
  {Sharpness of the {Berezinskii}-{Kosterlitz}-{Thouless} {Transition} in
  {Disordered} {NbN} {Films}},}\ }\href {\doibase
  10.1103/PhysRevLett.131.186002} {\bibfield  {journal} {\bibinfo  {journal}
  {Physical Review Letters}\ }\textbf {\bibinfo {volume} {131}},\ \bibinfo
  {pages} {186002} (\bibinfo {year} {2023})},\ \bibinfo {note} {publisher:
  American Physical Society}\BibitemShut {NoStop}%
\bibitem [{\citenamefont {Mondal}\ \emph {et~al.}(2011)\citenamefont {Mondal},
  \citenamefont {Kamlapure}, \citenamefont {Chand}, \citenamefont {Saraswat},
  \citenamefont {Kumar}, \citenamefont {Jesudasan}, \citenamefont {Benfatto},
  \citenamefont {Tripathi},\ and\ \citenamefont
  {Raychaudhuri}}]{mondal_phase_2011}%
  \BibitemOpen
  \bibfield  {author} {\bibinfo {author} {\bibfnamefont {M.}~\bibnamefont
  {Mondal}}, \bibinfo {author} {\bibfnamefont {A.}~\bibnamefont {Kamlapure}},
  \bibinfo {author} {\bibfnamefont {M.}~\bibnamefont {Chand}}, \bibinfo
  {author} {\bibfnamefont {G.}~\bibnamefont {Saraswat}}, \bibinfo {author}
  {\bibfnamefont {S.}~\bibnamefont {Kumar}}, \bibinfo {author} {\bibfnamefont
  {J.}~\bibnamefont {Jesudasan}}, \bibinfo {author} {\bibfnamefont
  {L.}~\bibnamefont {Benfatto}}, \bibinfo {author} {\bibfnamefont
  {V.}~\bibnamefont {Tripathi}}, \ and\ \bibinfo {author} {\bibfnamefont
  {P.}~\bibnamefont {Raychaudhuri}},\ }\bibfield  {title} {\enquote {\bibinfo
  {title} {Phase {Fluctuations} in a {Strongly} {Disordered} s-{Wave} {NbN}
  {Superconductor} {Close} to the {Metal}-{Insulator} {Transition}},}\ }\href
  {\doibase 10.1103/PhysRevLett.106.047001} {\bibfield  {journal} {\bibinfo
  {journal} {Physical Review Letters}\ }\textbf {\bibinfo {volume} {106}},\
  \bibinfo {pages} {047001} (\bibinfo {year} {2011})},\ \bibinfo {note}
  {publisher: American Physical Society}\BibitemShut {NoStop}%
\bibitem [{\citenamefont {Jaeger}\ \emph {et~al.}(1986)\citenamefont {Jaeger},
  \citenamefont {Haviland}, \citenamefont {Goldman},\ and\ \citenamefont
  {Orr}}]{jaeger_threshold_1986}%
  \BibitemOpen
  \bibfield  {author} {\bibinfo {author} {\bibfnamefont {H.~M.}\ \bibnamefont
  {Jaeger}}, \bibinfo {author} {\bibfnamefont {D.~B.}\ \bibnamefont
  {Haviland}}, \bibinfo {author} {\bibfnamefont {A.~M.}\ \bibnamefont
  {Goldman}}, \ and\ \bibinfo {author} {\bibfnamefont {B.~G.}\ \bibnamefont
  {Orr}},\ }\bibfield  {title} {\enquote {\bibinfo {title} {Threshold for
  superconductivity in ultrathin amorphous gallium films},}\ }\href {\doibase
  10.1103/PhysRevB.34.4920} {\bibfield  {journal} {\bibinfo  {journal}
  {Physical Review B}\ }\textbf {\bibinfo {volume} {34}},\ \bibinfo {pages}
  {4920--4923} (\bibinfo {year} {1986})},\ \bibinfo {note} {publisher: American
  Physical Society}\BibitemShut {NoStop}%
\bibitem [{\citenamefont {Sac\'ep\'e}\ \emph {et~al.}(2011)\citenamefont
  {Sac\'ep\'e}, \citenamefont {Dubouchet}, \citenamefont {Chapelier},
  \citenamefont {Sanquer}, \citenamefont {Ovadia}, \citenamefont {Shahar},
  \citenamefont {Feigel{’}man},\ and\ \citenamefont
  {Ioffe}}]{sacepe_localization_2011}%
  \BibitemOpen
  \bibfield  {author} {\bibinfo {author} {\bibfnamefont {B.}~\bibnamefont
  {Sac\'ep\'e}}, \bibinfo {author} {\bibfnamefont {T.}~\bibnamefont {Dubouchet}},
  \bibinfo {author} {\bibfnamefont {C.}~\bibnamefont {Chapelier}}, \bibinfo
  {author} {\bibfnamefont {M.}~\bibnamefont {Sanquer}}, \bibinfo {author}
  {\bibfnamefont {M.}~\bibnamefont {Ovadia}}, \bibinfo {author} {\bibfnamefont
  {D.}~\bibnamefont {Shahar}}, \bibinfo {author} {\bibfnamefont
  {M.}~\bibnamefont {Feigel{’}man}}, \ and\ \bibinfo {author} {\bibfnamefont
  {L.}~\bibnamefont {Ioffe}},\ }\bibfield  {title} {{\enquote {\bibinfo {title} {Localization of preformed {Cooper} pairs in
  disordered superconductors},}\ }}\href {\doibase 10.1038/nphys1892}
  {\bibfield  {journal} {\bibinfo  {journal} {Nature Physics}\ }\textbf
  {\bibinfo {volume} {7}},\ \bibinfo {pages} {239--244} (\bibinfo {year}
  {2011})},\ \bibinfo {note} {number: 3 Publisher: Nature Publishing
  Group}\BibitemShut {NoStop}%
\bibitem [{\citenamefont {Maleeva}\ \emph {et~al.}(2018)\citenamefont
  {Maleeva}, \citenamefont {Gr\"unhaupt}, \citenamefont {Klein}, \citenamefont
  {Levy-Bertrand}, \citenamefont {Dupre}, \citenamefont {Calvo}, \citenamefont
  {Valenti}, \citenamefont {Winkel}, \citenamefont {Friedrich}, \citenamefont
  {Wernsdorfer}, \citenamefont {Ustinov}, \citenamefont {Rotzinger},
  \citenamefont {Monfardini}, \citenamefont {Fistul},\ and\ \citenamefont
  {Pop}}]{maleeva_circuit_2018}%
  \BibitemOpen
  \bibfield  {author} {\bibinfo {author} {\bibfnamefont {N.}~\bibnamefont
  {Maleeva}}, \bibinfo {author} {\bibfnamefont {L.}~\bibnamefont {Gr\"unhaupt}},
  \bibinfo {author} {\bibfnamefont {T.}~\bibnamefont {Klein}}, \bibinfo
  {author} {\bibfnamefont {F.}~\bibnamefont {Levy-Bertrand}}, \bibinfo {author}
  {\bibfnamefont {O.}~\bibnamefont {Dupre}}, \bibinfo {author} {\bibfnamefont
  {M.}~\bibnamefont {Calvo}}, \bibinfo {author} {\bibfnamefont
  {F.}~\bibnamefont {Valenti}}, \bibinfo {author} {\bibfnamefont
  {P.}~\bibnamefont {Winkel}}, \bibinfo {author} {\bibfnamefont
  {F.}~\bibnamefont {Friedrich}}, \bibinfo {author} {\bibfnamefont
  {W.}~\bibnamefont {Wernsdorfer}}, \bibinfo {author} {\bibfnamefont {A.~V.}\
  \bibnamefont {Ustinov}}, \bibinfo {author} {\bibfnamefont {H.}~\bibnamefont
  {Rotzinger}}, \bibinfo {author} {\bibfnamefont {A.}~\bibnamefont
  {Monfardini}}, \bibinfo {author} {\bibfnamefont {M.~V.}\ \bibnamefont
  {Fistul}}, \ and\ \bibinfo {author} {\bibfnamefont {I.~M.}\ \bibnamefont
  {Pop}},\ }\bibfield  {title} {{\enquote {\bibinfo {title}
  {Circuit quantum electrodynamics of granular aluminum resonators},}\ }}\href
  {\doibase 10.1038/s41467-018-06386-9} {\bibfield  {journal} {\bibinfo
  {journal} {Nature Communications}\ }\textbf {\bibinfo {volume} {9}},\
  \bibinfo {pages} {3889} (\bibinfo {year} {2018})},\ \bibinfo {note} {number:
  1 Publisher: Nature Publishing Group}\BibitemShut {NoStop}%
\bibitem [{\citenamefont {Gr\"unhaupt}\ \emph {et~al.}(2019)\citenamefont
  {Gr\"unhaupt}, \citenamefont {Spiecker}, \citenamefont {Gusenkova},
  \citenamefont {Maleeva}, \citenamefont {Skacel}, \citenamefont {Takmakov},
  \citenamefont {Valenti}, \citenamefont {Winkel}, \citenamefont {Rotzinger},
  \citenamefont {Wernsdorfer}, \citenamefont {Ustinov},\ and\ \citenamefont
  {Pop}}]{grunhaupt_granular_2019-1}%
  \BibitemOpen
  \bibfield  {author} {\bibinfo {author} {\bibfnamefont {L.}~\bibnamefont
  {Gr\"unhaupt}}, \bibinfo {author} {\bibfnamefont {M.}~\bibnamefont
  {Spiecker}}, \bibinfo {author} {\bibfnamefont {D.}~\bibnamefont {Gusenkova}},
  \bibinfo {author} {\bibfnamefont {N.}~\bibnamefont {Maleeva}}, \bibinfo
  {author} {\bibfnamefont {S.~T.}\ \bibnamefont {Skacel}}, \bibinfo {author}
  {\bibfnamefont {I.}~\bibnamefont {Takmakov}}, \bibinfo {author}
  {\bibfnamefont {F.}~\bibnamefont {Valenti}}, \bibinfo {author} {\bibfnamefont
  {P.}~\bibnamefont {Winkel}}, \bibinfo {author} {\bibfnamefont
  {H.}~\bibnamefont {Rotzinger}}, \bibinfo {author} {\bibfnamefont
  {W.}~\bibnamefont {Wernsdorfer}}, \bibinfo {author} {\bibfnamefont {A.~V.}\
  \bibnamefont {Ustinov}}, \ and\ \bibinfo {author} {\bibfnamefont {I.~M.}\
  \bibnamefont {Pop}},\ }\bibfield  {title} {{\enquote
  {\bibinfo {title} {Granular aluminium as a superconducting material for
  high-impedance quantum circuits},}\ }}\href {\doibase
  10.1038/s41563-019-0350-3} {\bibfield  {journal} {\bibinfo  {journal} {Nature
  Materials}\ }\textbf {\bibinfo {volume} {18}},\ \bibinfo {pages} {816--819}
  (\bibinfo {year} {2019})},\ \bibinfo {note} {number: 8 Publisher: Nature
  Publishing Group}\BibitemShut {NoStop}%
\bibitem [{\citenamefont {Samkharadze}\ \emph {et~al.}(2016)\citenamefont
  {Samkharadze}, \citenamefont {Bruno}, \citenamefont {Scarlino}, \citenamefont
  {Zheng}, \citenamefont {DiVincenzo}, \citenamefont {DiCarlo},\ and\
  \citenamefont {Vandersypen}}]{samkharadze_high-kinetic-inductance_2016}%
  \BibitemOpen
  \bibfield  {author} {\bibinfo {author} {\bibfnamefont {N.}~\bibnamefont
  {Samkharadze}}, \bibinfo {author} {\bibfnamefont {A.}~\bibnamefont {Bruno}},
  \bibinfo {author} {\bibfnamefont {P.}~\bibnamefont {Scarlino}}, \bibinfo
  {author} {\bibfnamefont {G.}~\bibnamefont {Zheng}}, \bibinfo {author}
  {\bibfnamefont {D.}~\bibnamefont {DiVincenzo}}, \bibinfo {author}
  {\bibfnamefont {L.}~\bibnamefont {DiCarlo}}, \ and\ \bibinfo {author}
  {\bibfnamefont {L.}~\bibnamefont {Vandersypen}},\ }\bibfield  {title}
  {\enquote {\bibinfo {title} {High-{Kinetic}-{Inductance} {Superconducting}
  {Nanowire} {Resonators} for {Circuit} {QED} in a {Magnetic} {Field}},}\
  }\href {\doibase 10.1103/PhysRevApplied.5.044004} {\bibfield  {journal}
  {\bibinfo  {journal} {Physical Review Applied}\ }\textbf {\bibinfo {volume}
  {5}},\ \bibinfo {pages} {044004} (\bibinfo {year} {2016})},\ \bibinfo {note}
  {publisher: American Physical Society}\BibitemShut {NoStop}%
\bibitem [{\citenamefont {Lehtinen}, \citenamefont {Zakharov},\ and\
  \citenamefont {Arutyunov}(2012)}]{lehtinen_coulomb_2012}%
  \BibitemOpen
  \bibfield  {author} {\bibinfo {author} {\bibfnamefont {J.~S.}\ \bibnamefont
  {Lehtinen}}, \bibinfo {author} {\bibfnamefont {K.}~\bibnamefont {Zakharov}},
  \ and\ \bibinfo {author} {\bibfnamefont {K.~Y.}\ \bibnamefont {Arutyunov}},\
  }\bibfield  {title} {\enquote {\bibinfo {title} {Coulomb {Blockade} and
  {Bloch} {Oscillations} in {Superconducting} {Ti} {Nanowires}},}\ }\href
  {\doibase 10.1103/PhysRevLett.109.187001} {\bibfield  {journal} {\bibinfo
  {journal} {Physical Review Letters}\ }\textbf {\bibinfo {volume} {109}},\
  \bibinfo {pages} {187001} (\bibinfo {year} {2012})},\ \bibinfo {note}
  {publisher: American Physical Society}\BibitemShut {NoStop}%
\bibitem [{\citenamefont {Shaikhaidarov}\ \emph {et~al.}(2022)\citenamefont
  {Shaikhaidarov}, \citenamefont {Kim}, \citenamefont {Dunstan}, \citenamefont
  {Antonov}, \citenamefont {Linzen}, \citenamefont {Ziegler}, \citenamefont
  {Golubev}, \citenamefont {Antonov}, \citenamefont {Il{'}ichev},\ and\
  \citenamefont {Astafiev}}]{shaikhaidarov_quantized_2022}%
  \BibitemOpen
  \bibfield  {author} {\bibinfo {author} {\bibfnamefont {R.~S.}\ \bibnamefont
  {Shaikhaidarov}}, \bibinfo {author} {\bibfnamefont {K.~H.}\ \bibnamefont
  {Kim}}, \bibinfo {author} {\bibfnamefont {J.~W.}\ \bibnamefont {Dunstan}},
  \bibinfo {author} {\bibfnamefont {I.~V.}\ \bibnamefont {Antonov}}, \bibinfo
  {author} {\bibfnamefont {S.}~\bibnamefont {Linzen}}, \bibinfo {author}
  {\bibfnamefont {M.}~\bibnamefont {Ziegler}}, \bibinfo {author} {\bibfnamefont
  {D.~S.}\ \bibnamefont {Golubev}}, \bibinfo {author} {\bibfnamefont {V.~N.}\
  \bibnamefont {Antonov}}, \bibinfo {author} {\bibfnamefont {E.~V.}\
  \bibnamefont {Il{'}ichev}}, \ and\ \bibinfo {author} {\bibfnamefont {O.~V.}\
  \bibnamefont {Astafiev}},\ }\bibfield  {title} {{\enquote
  {\bibinfo {title} {Quantized current steps due to the a.c. coherent quantum
  phase-slip effect},}\ }}\href {\doibase 10.1038/s41586-022-04947-z}
  {\bibfield  {journal} {\bibinfo  {journal} {Nature}\ }\textbf {\bibinfo
  {volume} {608}},\ \bibinfo {pages} {45--49} (\bibinfo {year} {2022})},\
  \bibinfo {note} {number: 7921 Publisher: Nature Publishing Group}\BibitemShut
  {NoStop}%
\bibitem [{\citenamefont {Gol{'}tsman}\ \emph {et~al.}(2001)\citenamefont
  {Gol{'}tsman}, \citenamefont {Okunev}, \citenamefont {Chulkova},
  \citenamefont {Lipatov}, \citenamefont {Semenov}, \citenamefont {Smirnov},
  \citenamefont {Voronov}, \citenamefont {Dzardanov}, \citenamefont
  {Williams},\ and\ \citenamefont {Sobolewski}}]{goltsman_picosecond_2001}%
  \BibitemOpen
  \bibfield  {author} {\bibinfo {author} {\bibfnamefont {G.~N.}\ \bibnamefont
  {Gol{'}tsman}}, \bibinfo {author} {\bibfnamefont {O.}~\bibnamefont {Okunev}},
  \bibinfo {author} {\bibfnamefont {G.}~\bibnamefont {Chulkova}}, \bibinfo
  {author} {\bibfnamefont {A.}~\bibnamefont {Lipatov}}, \bibinfo {author}
  {\bibfnamefont {A.}~\bibnamefont {Semenov}}, \bibinfo {author} {\bibfnamefont
  {K.}~\bibnamefont {Smirnov}}, \bibinfo {author} {\bibfnamefont
  {B.}~\bibnamefont {Voronov}}, \bibinfo {author} {\bibfnamefont
  {A.}~\bibnamefont {Dzardanov}}, \bibinfo {author} {\bibfnamefont
  {C.}~\bibnamefont {Williams}}, \ and\ \bibinfo {author} {\bibfnamefont
  {R.}~\bibnamefont {Sobolewski}},\ }\bibfield  {title} {\enquote {\bibinfo
  {title} {Picosecond superconducting single-photon optical detector},}\ }\href
  {\doibase 10.1063/1.1388868} {\bibfield  {journal} {\bibinfo  {journal}
  {Applied Physics Letters}\ }\textbf {\bibinfo {volume} {79}},\ \bibinfo
  {pages} {705--707} (\bibinfo {year} {2001})},\ \bibinfo {note} {publisher:
  American Institute of Physics}\BibitemShut {NoStop}%
\bibitem [{\citenamefont {Esmaeil~Zadeh}\ \emph {et~al.}(2021)\citenamefont
  {Esmaeil~Zadeh}, \citenamefont {Chang}, \citenamefont {Los}, \citenamefont
  {Gyger}, \citenamefont {Elshaari}, \citenamefont {Steinhauer}, \citenamefont
  {Dorenbos},\ and\ \citenamefont
  {Zwiller}}]{esmaeil_zadeh_superconducting_2021}%
  \BibitemOpen
  \bibfield  {author} {\bibinfo {author} {\bibfnamefont {I.}~\bibnamefont
  {Esmaeil~Zadeh}}, \bibinfo {author} {\bibfnamefont {J.}~\bibnamefont
  {Chang}}, \bibinfo {author} {\bibfnamefont {J.~W.~N.}\ \bibnamefont {Los}},
  \bibinfo {author} {\bibfnamefont {S.}~\bibnamefont {Gyger}}, \bibinfo
  {author} {\bibfnamefont {A.~W.}\ \bibnamefont {Elshaari}}, \bibinfo {author}
  {\bibfnamefont {S.}~\bibnamefont {Steinhauer}}, \bibinfo {author}
  {\bibfnamefont {S.~N.}\ \bibnamefont {Dorenbos}}, \ and\ \bibinfo {author}
  {\bibfnamefont {V.}~\bibnamefont {Zwiller}},\ }\bibfield  {title} {\enquote
  {\bibinfo {title} {Superconducting nanowire single-photon detectors: {A}
  perspective on evolution, state-of-the-art, future developments, and
  applications},}\ }\href {\doibase 10.1063/5.0045990} {\bibfield  {journal}
  {\bibinfo  {journal} {Applied Physics Letters}\ }\textbf {\bibinfo {volume}
  {118}},\ \bibinfo {pages} {190502} (\bibinfo {year} {2021})},\ \bibinfo
  {note} {publisher: American Institute of Physics}\BibitemShut {NoStop}%
\bibitem [{\citenamefont {Gantmakher}\ and\ \citenamefont
  {Dolgopolov}(2010)}]{gantmakher_superconductorinsulator_2010}%
  \BibitemOpen
  \bibfield  {author} {\bibinfo {author} {\bibfnamefont {V.~F.}\ \bibnamefont
  {Gantmakher}}\ and\ \bibinfo {author} {\bibfnamefont {V.~T.}\ \bibnamefont
  {Dolgopolov}},\ }\bibfield  {title} {{\enquote {\bibinfo
  {title} {Superconductor–insulator quantum phase transition},}\ }}\href
  {\doibase 10.3367/UFNe.0180.201001a.0003} {\bibfield  {journal} {\bibinfo
  {journal} {Physics-Uspekhi}\ }\textbf {\bibinfo {volume} {53}},\ \bibinfo
  {pages} {1} (\bibinfo {year} {2010})},\ \bibinfo {note} {publisher: IOP
  Publishing}\BibitemShut {NoStop}%
\bibitem [{\citenamefont {Zou}\ \emph {et~al.}(2017)\citenamefont {Zou},
  \citenamefont {Chen}, \citenamefont {Zhang}, \citenamefont {Xiu},
  \citenamefont {Matsumura}, \citenamefont {Yang}, \citenamefont {Hong},\ and\
  \citenamefont {Zou}}]{zou_superconductivity_2017}%
  \BibitemOpen
  \bibfield  {author} {\bibinfo {author} {\bibfnamefont {Y.-C.}\ \bibnamefont
  {Zou}}, \bibinfo {author} {\bibfnamefont {Z.-G.}\ \bibnamefont {Chen}},
  \bibinfo {author} {\bibfnamefont {E.}~\bibnamefont {Zhang}}, \bibinfo
  {author} {\bibfnamefont {F.}~\bibnamefont {Xiu}}, \bibinfo {author}
  {\bibfnamefont {S.}~\bibnamefont {Matsumura}}, \bibinfo {author}
  {\bibfnamefont {L.}~\bibnamefont {Yang}}, \bibinfo {author} {\bibfnamefont
  {M.}~\bibnamefont {Hong}}, \ and\ \bibinfo {author} {\bibfnamefont
  {J.}~\bibnamefont {Zou}},\ }\bibfield  {title} {{\enquote
  {\bibinfo {title} {Superconductivity and magnetotransport of
  single-crystalline {NbSe2} nanoplates grown by chemical vapour deposition},}\
  }}\href {\doibase 10.1039/C7NR06617A} {\bibfield  {journal} {\bibinfo
  {journal} {Nanoscale}\ }\textbf {\bibinfo {volume} {9}},\ \bibinfo {pages}
  {16591--16595} (\bibinfo {year} {2017})},\ \bibinfo {note} {publisher: Royal
  Society of Chemistry}\BibitemShut {NoStop}%
\bibitem [{\citenamefont {Cao}\ \emph {et~al.}(2018)\citenamefont {Cao},
  \citenamefont {Fatemi}, \citenamefont {Fang}, \citenamefont {Watanabe},
  \citenamefont {Taniguchi}, \citenamefont {Kaxiras},\ and\ \citenamefont
  {Jarillo-Herrero}}]{cao_unconventional_2018}%
  \BibitemOpen
  \bibfield  {author} {\bibinfo {author} {\bibfnamefont {Y.}~\bibnamefont
  {Cao}}, \bibinfo {author} {\bibfnamefont {V.}~\bibnamefont {Fatemi}},
  \bibinfo {author} {\bibfnamefont {S.}~\bibnamefont {Fang}}, \bibinfo {author}
  {\bibfnamefont {K.}~\bibnamefont {Watanabe}}, \bibinfo {author}
  {\bibfnamefont {T.}~\bibnamefont {Taniguchi}}, \bibinfo {author}
  {\bibfnamefont {E.}~\bibnamefont {Kaxiras}}, \ and\ \bibinfo {author}
  {\bibfnamefont {P.}~\bibnamefont {Jarillo-Herrero}},\ }\bibfield  {title}
  {{\enquote {\bibinfo {title} {Unconventional
  superconductivity in magic-angle graphene superlattices},}\ }}\href {\doibase
  10.1038/nature26160} {\bibfield  {journal} {\bibinfo  {journal} {Nature}\
  }\textbf {\bibinfo {volume} {556}},\ \bibinfo {pages} {43--50} (\bibinfo
  {year} {2018})},\ \bibinfo {note} {number: 7699 Publisher: Nature Publishing
  Group}\BibitemShut {NoStop}%
\bibitem [{\citenamefont {Buckel}\ and\ \citenamefont
  {Hilsch}(1954)}]{buckel_einflus_1954}%
  \BibitemOpen
  \bibfield  {author} {\bibinfo {author} {\bibfnamefont {W.}~\bibnamefont
  {Buckel}}\ and\ \bibinfo {author} {\bibfnamefont {R.}~\bibnamefont
  {Hilsch}},\ }\bibfield  {title} {{\enquote {\bibinfo
  {title} {Einflu{\ss} der {Kondensation} bei tiefen {Temperaturen} auf den
  elektrischen {Widerstand} und die {Supraleitung} f\"ur verschiedene
  {Metalle}},}\ }}\href {\doibase 10.1007/BF01337903} {\bibfield  {journal}
  {\bibinfo  {journal} {Zeitschrift f\"ur Physik}\ }\textbf {\bibinfo {volume}
  {138}},\ \bibinfo {pages} {109--120} (\bibinfo {year} {1954})}\BibitemShut
  {NoStop}%
\bibitem [{\citenamefont {Kammerer}\ and\ \citenamefont
  {Strongin}(1965)}]{kammerer_superconductivity_1965}%
  \BibitemOpen
  \bibfield  {author} {\bibinfo {author} {\bibfnamefont {O.~F.}\ \bibnamefont
  {Kammerer}}\ and\ \bibinfo {author} {\bibfnamefont {M.}~\bibnamefont
  {Strongin}},\ }\bibfield  {title} {{\enquote {\bibinfo
  {title} {Superconductivity in tungsten films},}\ }}\href {\doibase
  10.1016/0031-9163(65)90496-8} {\bibfield  {journal} {\bibinfo  {journal}
  {Physics Letters}\ }\textbf {\bibinfo {volume} {17}},\ \bibinfo {pages}
  {224--225} (\bibinfo {year} {1965})}\BibitemShut {NoStop}%
\bibitem [{\citenamefont {Abeles}, \citenamefont {Cohen},\ and\ \citenamefont
  {Cullen}(1966)}]{abeles_enhancement_1966}%
  \BibitemOpen
  \bibfield  {author} {\bibinfo {author} {\bibfnamefont {B.}~\bibnamefont
  {Abeles}}, \bibinfo {author} {\bibfnamefont {R.~W.}\ \bibnamefont {Cohen}}, \
  and\ \bibinfo {author} {\bibfnamefont {G.~W.}\ \bibnamefont {Cullen}},\
  }\bibfield  {title} {\enquote {\bibinfo {title} {Enhancement of
  {Superconductivity} in {Metal} {Films}},}\ }\href {\doibase
  10.1103/PhysRevLett.17.632} {\bibfield  {journal} {\bibinfo  {journal}
  {Physical Review Letters}\ }\textbf {\bibinfo {volume} {17}},\ \bibinfo
  {pages} {632--634} (\bibinfo {year} {1966})},\ \bibinfo {note} {publisher:
  American Physical Society}\BibitemShut {NoStop}%
\bibitem [{\citenamefont {Strongin}\ \emph {et~al.}(1967)\citenamefont
  {Strongin}, \citenamefont {Kammerer}, \citenamefont {Douglass},\ and\
  \citenamefont {Cohen}}]{strongin_effect_1967}%
  \BibitemOpen
  \bibfield  {author} {\bibinfo {author} {\bibfnamefont {M.}~\bibnamefont
  {Strongin}}, \bibinfo {author} {\bibfnamefont {O.~F.}\ \bibnamefont
  {Kammerer}}, \bibinfo {author} {\bibfnamefont {D.~H.}\ \bibnamefont
  {Douglass}}, \ and\ \bibinfo {author} {\bibfnamefont {M.~H.}\ \bibnamefont
  {Cohen}},\ }\bibfield  {title} {\enquote {\bibinfo {title} {Effect of
  {Dielectric} and {High}-{Resistivity} {Barriers} on the {Superconducting}
  {Transition} {Temperature} of {Thin} {Films}},}\ }\href {\doibase
  10.1103/PhysRevLett.19.121} {\bibfield  {journal} {\bibinfo  {journal}
  {Physical Review Letters}\ }\textbf {\bibinfo {volume} {19}},\ \bibinfo
  {pages} {121--125} (\bibinfo {year} {1967})},\ \bibinfo {note} {publisher:
  American Physical Society}\BibitemShut {NoStop}%
\bibitem [{\citenamefont {Pettit}\ and\ \citenamefont
  {Silcox}(1976)}]{pettit_film_1976}%
  \BibitemOpen
  \bibfield  {author} {\bibinfo {author} {\bibfnamefont {R.~B.}\ \bibnamefont
  {Pettit}}\ and\ \bibinfo {author} {\bibfnamefont {J.}~\bibnamefont
  {Silcox}},\ }\bibfield  {title} {\enquote {\bibinfo {title} {Film structure
  and enhanced superconductivity in evaporated aluminum films},}\ }\href
  {\doibase 10.1103/PhysRevB.13.2865} {\bibfield  {journal} {\bibinfo
  {journal} {Physical Review B}\ }\textbf {\bibinfo {volume} {13}},\ \bibinfo
  {pages} {2865--2872} (\bibinfo {year} {1976})},\ \bibinfo {note} {publisher:
  American Physical Society}\BibitemShut {NoStop}%
\bibitem [{\citenamefont {Gr\"unhaupt}(2019)}]{grunhaupt_granular_2019}%
  \BibitemOpen
  \bibfield  {author} {\bibinfo {author} {\bibfnamefont {L.}~\bibnamefont
  {Gr\"unhaupt}},\ }\href {\doibase 10.5445/KSP/1000097320} {{\emph {\bibinfo {title} {Granular aluminium superinductors}}}}\ (\bibinfo
   {publisher} {KIT Scientific Publishing},\ \bibinfo {year} {2019})\ \bibinfo
  {note} {publication Title: KIT Scientific Publishing}\BibitemShut {NoStop}%
\bibitem [{\citenamefont {Bose}\ \emph {et~al.}(2005)\citenamefont {Bose},
  \citenamefont {Raychaudhuri}, \citenamefont {Banerjee}, \citenamefont
  {Vasa},\ and\ \citenamefont {Ayyub}}]{bose_mechanism_2005}%
  \BibitemOpen
  \bibfield  {author} {\bibinfo {author} {\bibfnamefont {S.}~\bibnamefont
  {Bose}}, \bibinfo {author} {\bibfnamefont {P.}~\bibnamefont {Raychaudhuri}},
  \bibinfo {author} {\bibfnamefont {R.}~\bibnamefont {Banerjee}}, \bibinfo
  {author} {\bibfnamefont {P.}~\bibnamefont {Vasa}}, \ and\ \bibinfo {author}
  {\bibfnamefont {P.}~\bibnamefont {Ayyub}},\ }\bibfield  {title} {\enquote
  {\bibinfo {title} {Mechanism of the {Size} {Dependence} of the
  {Superconducting} {Transition} of {Nanostructured} {Nb}},}\ }\href {\doibase
  10.1103/PhysRevLett.95.147003} {\bibfield  {journal} {\bibinfo  {journal}
  {Physical Review Letters}\ }\textbf {\bibinfo {volume} {95}},\ \bibinfo
  {pages} {147003} (\bibinfo {year} {2005})},\ \bibinfo {note} {publisher:
  American Physical Society}\BibitemShut {NoStop}%
\bibitem [{\citenamefont {Tanatar}\ \emph {et~al.}(2022)\citenamefont
  {Tanatar}, \citenamefont {Torsello}, \citenamefont {Joshi}, \citenamefont
  {Ghimire}, \citenamefont {Kopas}, \citenamefont {Marshall}, \citenamefont
  {Mutus}, \citenamefont {Ghigo}, \citenamefont {Zarea}, \citenamefont
  {Sauls},\ and\ \citenamefont {Prozorov}}]{tanatar_anisotropic_2022}%
  \BibitemOpen
  \bibfield  {author} {\bibinfo {author} {\bibfnamefont {M.~A.}\ \bibnamefont
  {Tanatar}}, \bibinfo {author} {\bibfnamefont {D.}~\bibnamefont {Torsello}},
  \bibinfo {author} {\bibfnamefont {K.~R.}\ \bibnamefont {Joshi}}, \bibinfo
  {author} {\bibfnamefont {S.}~\bibnamefont {Ghimire}}, \bibinfo {author}
  {\bibfnamefont {C.~J.}\ \bibnamefont {Kopas}}, \bibinfo {author}
  {\bibfnamefont {J.}~\bibnamefont {Marshall}}, \bibinfo {author}
  {\bibfnamefont {J.~Y.}\ \bibnamefont {Mutus}}, \bibinfo {author}
  {\bibfnamefont {G.}~\bibnamefont {Ghigo}}, \bibinfo {author} {\bibfnamefont
  {M.}~\bibnamefont {Zarea}}, \bibinfo {author} {\bibfnamefont {J.~A.}\
  \bibnamefont {Sauls}}, \ and\ \bibinfo {author} {\bibfnamefont
  {R.}~\bibnamefont {Prozorov}},\ }\bibfield  {title} {\enquote {\bibinfo
  {title} {Anisotropic superconductivity of niobium based on its response to
  nonmagnetic disorder},}\ }\href {\doibase 10.1103/PhysRevB.106.224511}
  {\bibfield  {journal} {\bibinfo  {journal} {Physical Review B}\ }\textbf
  {\bibinfo {volume} {106}},\ \bibinfo {pages} {224511} (\bibinfo {year}
  {2022})},\ \bibinfo {note} {publisher: American Physical Society}\BibitemShut
  {NoStop}%
\bibitem [{\citenamefont {Sweedler}, \citenamefont {Schweitzer},\ and\
  \citenamefont {Webb}(1974)}]{sweedler_atomic_1974}%
  \BibitemOpen
  \bibfield  {author} {\bibinfo {author} {\bibfnamefont {A.~R.}\ \bibnamefont
  {Sweedler}}, \bibinfo {author} {\bibfnamefont {D.~G.}\ \bibnamefont
  {Schweitzer}}, \ and\ \bibinfo {author} {\bibfnamefont {G.~W.}\ \bibnamefont
  {Webb}},\ }\bibfield  {title} {\enquote {\bibinfo {title} {Atomic {Ordering}
  and {Superconductivity} in {High}-{Tc} {A}-15 {Compounds}},}\ }\href
  {\doibase 10.1103/PhysRevLett.33.168} {\bibfield  {journal} {\bibinfo
  {journal} {Physical Review Letters}\ }\textbf {\bibinfo {volume} {33}},\
  \bibinfo {pages} {168--172} (\bibinfo {year} {1974})},\ \bibinfo {note}
  {publisher: American Physical Society}\BibitemShut {NoStop}%
\bibitem [{\citenamefont {Lita}\ \emph {et~al.}(2005)\citenamefont {Lita},
  \citenamefont {Rosenberg}, \citenamefont {Nam}, \citenamefont {Miller},
  \citenamefont {Balzar}, \citenamefont {Kaatz},\ and\ \citenamefont
  {Schwall}}]{lita_tuning_2005}%
  \BibitemOpen
  \bibfield  {author} {\bibinfo {author} {\bibfnamefont {A.~E.}\ \bibnamefont
  {Lita}}, \bibinfo {author} {\bibfnamefont {D.}~\bibnamefont {Rosenberg}},
  \bibinfo {author} {\bibfnamefont {S.}~\bibnamefont {Nam}}, \bibinfo {author}
  {\bibfnamefont {A.~J.}\ \bibnamefont {Miller}}, \bibinfo {author}
  {\bibfnamefont {D.}~\bibnamefont {Balzar}}, \bibinfo {author} {\bibfnamefont
  {L.~M.}\ \bibnamefont {Kaatz}}, \ and\ \bibinfo {author} {\bibfnamefont
  {R.~E.}\ \bibnamefont {Schwall}},\ }\bibfield  {title} {\enquote {\bibinfo
  {title} {Tuning of tungsten thin film superconducting transition temperature
  for fabrication of photon number resolving detectors},}\ }\href {\doibase
  10.1109/TASC.2005.849033} {\bibfield  {journal} {\bibinfo  {journal} {IEEE
  Transactions on Applied Superconductivity}\ }\textbf {\bibinfo {volume}
  {15}},\ \bibinfo {pages} {3528--3531} (\bibinfo {year} {2005})},\ \bibinfo
  {note} {conference Name: IEEE Transactions on Applied
  Superconductivity}\BibitemShut {NoStop}%
\bibitem [{\citenamefont {Osofsky}\ \emph {et~al.}(2001)\citenamefont
  {Osofsky}, \citenamefont {Soulen}, \citenamefont {Claassen}, \citenamefont
  {Trotter}, \citenamefont {Kim},\ and\ \citenamefont
  {Horwitz}}]{osofsky_new_2001}%
  \BibitemOpen
  \bibfield  {author} {\bibinfo {author} {\bibfnamefont {M.~S.}\ \bibnamefont
  {Osofsky}}, \bibinfo {author} {\bibfnamefont {R.~J.}\ \bibnamefont {Soulen}},
  \bibinfo {author} {\bibfnamefont {J.~H.}\ \bibnamefont {Claassen}}, \bibinfo
  {author} {\bibfnamefont {G.}~\bibnamefont {Trotter}}, \bibinfo {author}
  {\bibfnamefont {H.}~\bibnamefont {Kim}}, \ and\ \bibinfo {author}
  {\bibfnamefont {J.~S.}\ \bibnamefont {Horwitz}},\ }\bibfield  {title}
  {{\enquote {\bibinfo {title} {New {Insight} into
  {Enhanced} {Superconductivity} in {Metals} near the {Metal}-{Insulator}
  {Transition}},}\ }}\href {\doibase 10.1103/PhysRevLett.87.197004} {\bibfield
  {journal} {\bibinfo  {journal} {Physical Review Letters}\ }\textbf {\bibinfo
  {volume} {87}},\ \bibinfo {pages} {197004} (\bibinfo {year}
  {2001})}\BibitemShut {NoStop}%
\bibitem [{\citenamefont {Bosworth}\ \emph {et~al.}(2015)\citenamefont
  {Bosworth}, \citenamefont {Sahonta}, \citenamefont {Hadfield},\ and\
  \citenamefont {Barber}}]{bosworth_amorphous_2015}%
  \BibitemOpen
  \bibfield  {author} {\bibinfo {author} {\bibfnamefont {D.}~\bibnamefont
  {Bosworth}}, \bibinfo {author} {\bibfnamefont {S.-L.}\ \bibnamefont
  {Sahonta}}, \bibinfo {author} {\bibfnamefont {R.~H.}\ \bibnamefont
  {Hadfield}}, \ and\ \bibinfo {author} {\bibfnamefont {Z.~H.}\ \bibnamefont
  {Barber}},\ }\bibfield  {title} {{\enquote {\bibinfo
  {title} {Amorphous molybdenum silicon superconducting thin films},}\ }}\href
  {\doibase 10.1063/1.4928285} {\bibfield  {journal} {\bibinfo  {journal} {AIP
  Advances}\ }\textbf {\bibinfo {volume} {5}},\ \bibinfo {pages} {087106}
  (\bibinfo {year} {2015})}\BibitemShut {NoStop}%
\bibitem [{\citenamefont {Banerjee}\ \emph {et~al.}(2017)\citenamefont
  {Banerjee}, \citenamefont {Baker}, \citenamefont {Doye}, \citenamefont
  {Nord}, \citenamefont {Heath}, \citenamefont {Erotokritou}, \citenamefont
  {Bosworth}, \citenamefont {Barber}, \citenamefont {MacLaren},\ and\
  \citenamefont {Hadfield}}]{banerjee_characterisation_2017}%
  \BibitemOpen
  \bibfield  {author} {\bibinfo {author} {\bibfnamefont {A.}~\bibnamefont
  {Banerjee}}, \bibinfo {author} {\bibfnamefont {L.~J.}\ \bibnamefont {Baker}},
  \bibinfo {author} {\bibfnamefont {A.}~\bibnamefont {Doye}}, \bibinfo {author}
  {\bibfnamefont {M.}~\bibnamefont {Nord}}, \bibinfo {author} {\bibfnamefont
  {R.~M.}\ \bibnamefont {Heath}}, \bibinfo {author} {\bibfnamefont
  {K.}~\bibnamefont {Erotokritou}}, \bibinfo {author} {\bibfnamefont
  {D.}~\bibnamefont {Bosworth}}, \bibinfo {author} {\bibfnamefont {Z.~H.}\
  \bibnamefont {Barber}}, \bibinfo {author} {\bibfnamefont {I.}~\bibnamefont
  {MacLaren}}, \ and\ \bibinfo {author} {\bibfnamefont {R.~H.}\ \bibnamefont
  {Hadfield}},\ }\bibfield  {title} {{\enquote {\bibinfo
  {title} {Characterisation of amorphous molybdenum silicide ({MoSi})
  superconducting thin films and nanowires},}\ }}\href {\doibase
  10.1088/1361-6668/aa76d8} {\bibfield  {journal} {\bibinfo  {journal}
  {Superconductor Science and Technology}\ }\textbf {\bibinfo {volume} {30}},\
  \bibinfo {pages} {084010} (\bibinfo {year} {2017})}\BibitemShut {NoStop}%
\bibitem [{\citenamefont {Zhang}\ \emph {et~al.}(2021)\citenamefont {Zhang},
  \citenamefont {Charaev}, \citenamefont {Liu}, \citenamefont {Zhou},
  \citenamefont {Zhu}, \citenamefont {Berggren},\ and\ \citenamefont
  {Schilling}}]{zhang_physical_2021}%
  \BibitemOpen
  \bibfield  {author} {\bibinfo {author} {\bibfnamefont {X.}~\bibnamefont
  {Zhang}}, \bibinfo {author} {\bibfnamefont {I.}~\bibnamefont {Charaev}},
  \bibinfo {author} {\bibfnamefont {H.}~\bibnamefont {Liu}}, \bibinfo {author}
  {\bibfnamefont {T.~X.}\ \bibnamefont {Zhou}}, \bibinfo {author}
  {\bibfnamefont {D.}~\bibnamefont {Zhu}}, \bibinfo {author} {\bibfnamefont
  {K.~K.}\ \bibnamefont {Berggren}}, \ and\ \bibinfo {author} {\bibfnamefont
  {A.}~\bibnamefont {Schilling}},\ }\bibfield  {title} {{\enquote {\bibinfo {title} {Physical properties of amorphous molybdenum
  silicide films for single-photon detectors},}\ }}\href {\doibase
  10.1088/1361-6668/ac1524} {\bibfield  {journal} {\bibinfo  {journal}
  {Superconductor Science and Technology}\ }\textbf {\bibinfo {volume} {34}},\
  \bibinfo {pages} {095003} (\bibinfo {year} {2021})}\BibitemShut {NoStop}%
\bibitem [{\citenamefont {T\"ut\"unc\"u}, \citenamefont {Ba\u{g}cı},\ and\
  \citenamefont {Srivastava}(2010)}]{tutuncu_electronic_2010}%
  \BibitemOpen
  \bibfield  {author} {\bibinfo {author} {\bibfnamefont {H.~M.}\ \bibnamefont
  {T\"ut\"unc\"u}}, \bibinfo {author} {\bibfnamefont {S.}~\bibnamefont {Ba\u{g}cı}},
  \ and\ \bibinfo {author} {\bibfnamefont {G.~P.}\ \bibnamefont {Srivastava}},\
  }\bibfield  {title} {\enquote {\bibinfo {title} {Electronic structure,
  phonons, and electron-phonon interaction in
  Mo3Si},}\
  }\href {\doibase 10.1103/PhysRevB.82.214510} {\bibfield  {journal} {\bibinfo
  {journal} {Physical Review B}\ }\textbf {\bibinfo {volume} {82}},\ \bibinfo
  {pages} {214510} (\bibinfo {year} {2010})},\ \bibinfo {note} {publisher:
  American Physical Society}\BibitemShut {NoStop}%
\bibitem [{\citenamefont {Tsaur}, \citenamefont {Liau},\ and\ \citenamefont
  {Mayer}(1979)}]{tsaur_ionbeaminduced_1979}%
  \BibitemOpen
  \bibfield  {author} {\bibinfo {author} {\bibfnamefont {B.~Y.}\ \bibnamefont
  {Tsaur}}, \bibinfo {author} {\bibfnamefont {Z.~L.}\ \bibnamefont {Liau}}, \
  and\ \bibinfo {author} {\bibfnamefont {J.~W.}\ \bibnamefont {Mayer}},\
  }\bibfield  {title} {{\enquote {\bibinfo {title}
  {Ion‐beam‐induced silicide formation},}\ }}\href {\doibase
  10.1063/1.90716} {\bibfield  {journal} {\bibinfo  {journal} {Applied Physics
  Letters}\ }\textbf {\bibinfo {volume} {34}},\ \bibinfo {pages} {168--170}
  (\bibinfo {year} {1979})}\BibitemShut {NoStop}%
\bibitem [{\citenamefont {Lehtinen}\ \emph {et~al.}(2017)\citenamefont
  {Lehtinen}, \citenamefont {Kemppinen}, \citenamefont {Mykk\"anen},
  \citenamefont {Prunnila},\ and\ \citenamefont
  {Manninen}}]{lehtinen_superconducting_2017}%
  \BibitemOpen
  \bibfield  {author} {\bibinfo {author} {\bibfnamefont {J.~S.}\ \bibnamefont
  {Lehtinen}}, \bibinfo {author} {\bibfnamefont {A.}~\bibnamefont {Kemppinen}},
  \bibinfo {author} {\bibfnamefont {E.}~\bibnamefont {Mykk\"anen}}, \bibinfo
  {author} {\bibfnamefont {M.}~\bibnamefont {Prunnila}}, \ and\ \bibinfo
  {author} {\bibfnamefont {A.~J.}\ \bibnamefont {Manninen}},\ }\bibfield
  {title} {{\enquote {\bibinfo {title} {Superconducting
  {MoSi} nanowires},}\ }}\href {\doibase 10.1088/1361-6668/aa954b} {\bibfield
  {journal} {\bibinfo  {journal} {Superconductor Science and Technology}\
  }\textbf {\bibinfo {volume} {31}},\ \bibinfo {pages} {015002} (\bibinfo
  {year} {2017})},\ \bibinfo {note} {publisher: IOP Publishing}\BibitemShut
  {NoStop}%
\bibitem [{\citenamefont {Juang}\ \emph {et~al.}(1988)\citenamefont {Juang},
  \citenamefont {Rudman}, \citenamefont {Talvacchio},\ and\ \citenamefont
  {Van~Dover}}]{juang_effects_1988}%
  \BibitemOpen
  \bibfield  {author} {\bibinfo {author} {\bibfnamefont {J.~Y.}\ \bibnamefont
  {Juang}}, \bibinfo {author} {\bibfnamefont {D.~A.}\ \bibnamefont {Rudman}},
  \bibinfo {author} {\bibfnamefont {J.}~\bibnamefont {Talvacchio}}, \ and\
  \bibinfo {author} {\bibfnamefont {R.~B.}\ \bibnamefont {Van~Dover}},\
  }\bibfield  {title} {{\enquote {\bibinfo {title} {Effects
  of ion irradiation on the normal state and superconducting properties of
  {NbN} thin films},}\ }}\href {\doibase 10.1103/PhysRevB.38.2354} {\bibfield
  {journal} {\bibinfo  {journal} {Physical Review B}\ }\textbf {\bibinfo
  {volume} {38}},\ \bibinfo {pages} {2354--2361} (\bibinfo {year}
  {1988})}\BibitemShut {NoStop}%
\bibitem [{\citenamefont {Martinez}\ \emph {et~al.}(2020)\citenamefont
  {Martinez}, \citenamefont {Buckley}, \citenamefont {Charaev}, \citenamefont
  {Dane}, \citenamefont {Dow},\ and\ \citenamefont
  {Berggren}}]{martinez_superconducting_2020}%
  \BibitemOpen
  \bibfield  {author} {\bibinfo {author} {\bibfnamefont {G.~D.}\ \bibnamefont
  {Martinez}}, \bibinfo {author} {\bibfnamefont {D.}~\bibnamefont {Buckley}},
  \bibinfo {author} {\bibfnamefont {I.}~\bibnamefont {Charaev}}, \bibinfo
  {author} {\bibfnamefont {A.}~\bibnamefont {Dane}}, \bibinfo {author}
  {\bibfnamefont {D.~E.}\ \bibnamefont {Dow}}, \ and\ \bibinfo {author}
  {\bibfnamefont {K.~K.}\ \bibnamefont {Berggren}},\ }\href {\doibase
  10.48550/arXiv.2003.02898} {\enquote {\bibinfo {title} {Superconducting
  {Nanowire} {Fabrication} on {Niobium} {Nitride} using {Helium} {Ion}
  {Irradiation}},}\ } (\bibinfo {year} {2020}),\ \bibinfo {note}
  {arXiv:2003.02898}\BibitemShut {NoStop}%
\bibitem [{\citenamefont {Zhang}\ \emph {et~al.}(2019)\citenamefont {Zhang},
  \citenamefont {Jia}, \citenamefont {You}, \citenamefont {Ou}, \citenamefont
  {Huang}, \citenamefont {Zhang}, \citenamefont {Li}, \citenamefont {Wang},\
  and\ \citenamefont {Xie}}]{zhang_saturating_2019}%
  \BibitemOpen
  \bibfield  {author} {\bibinfo {author} {\bibfnamefont {W.}~\bibnamefont
  {Zhang}}, \bibinfo {author} {\bibfnamefont {Q.}~\bibnamefont {Jia}}, \bibinfo
  {author} {\bibfnamefont {L.}~\bibnamefont {You}}, \bibinfo {author}
  {\bibfnamefont {X.}~\bibnamefont {Ou}}, \bibinfo {author} {\bibfnamefont
  {H.}~\bibnamefont {Huang}}, \bibinfo {author} {\bibfnamefont
  {L.}~\bibnamefont {Zhang}}, \bibinfo {author} {\bibfnamefont
  {H.}~\bibnamefont {Li}}, \bibinfo {author} {\bibfnamefont {Z.}~\bibnamefont
  {Wang}}, \ and\ \bibinfo {author} {\bibfnamefont {X.}~\bibnamefont {Xie}},\
  }\bibfield  {title} {\enquote {\bibinfo {title} {Saturating {Intrinsic}
  {Detection} {Efficiency} of {Superconducting} {Nanowire} {Single}-{Photon}
  {Detectors} via {Defect} {Engineering}},}\ }\href {\doibase
  10.1103/PhysRevApplied.12.044040} {\bibfield  {journal} {\bibinfo  {journal}
  {Physical Review Applied}\ }\textbf {\bibinfo {volume} {12}},\ \bibinfo
  {pages} {044040} (\bibinfo {year} {2019})},\ \bibinfo {note} {publisher:
  American Physical Society}\BibitemShut {NoStop}%
\bibitem [{\citenamefont {Ruhtinas}\ and\ \citenamefont
  {Maasilta}(2023)}]{ruhtinas_highly_2023}%
  \BibitemOpen
  \bibfield  {author} {\bibinfo {author} {\bibfnamefont {A.}~\bibnamefont
  {Ruhtinas}}\ and\ \bibinfo {author} {\bibfnamefont {I.~J.}\ \bibnamefont
  {Maasilta}},\ }\href {\doibase 10.48550/arXiv.2303.17348} {\enquote {\bibinfo
  {title} {Highly tunable {NbTiN} {Josephson} junctions fabricated with focused
  helium ion beam},}\ } (\bibinfo {year} {2023}),\ \bibinfo {note}
  {arXiv:2303.17348}\BibitemShut {NoStop}%
\bibitem [{\citenamefont {Yabuno}\ \emph {et~al.}(2023)\citenamefont {Yabuno},
  \citenamefont {China}, \citenamefont {Terai},\ and\ \citenamefont
  {Miki}}]{yabuno_superconducting_2023}%
  \BibitemOpen
  \bibfield  {author} {\bibinfo {author} {\bibfnamefont {M.}~\bibnamefont
  {Yabuno}}, \bibinfo {author} {\bibfnamefont {F.}~\bibnamefont {China}},
  \bibinfo {author} {\bibfnamefont {H.}~\bibnamefont {Terai}}, \ and\ \bibinfo
  {author} {\bibfnamefont {S.}~\bibnamefont {Miki}},\ }\bibfield  {title}
  {{\enquote {\bibinfo {title} {Superconducting wide strip
  photon detector with high critical current bank structure},}\ }}\href
  {\doibase 10.1364/OPTICAQ.497675} {\bibfield  {journal} {\bibinfo  {journal}
  {Optica Quantum}\ }\textbf {\bibinfo {volume} {1}},\ \bibinfo {pages}
  {26--34} (\bibinfo {year} {2023})},\ \bibinfo {note} {publisher: Optica
  Publishing Group}\BibitemShut {NoStop}%
\bibitem [{\citenamefont {Mykk\"anen}\ \emph {et~al.}(2020)\citenamefont
  {Mykk\"anen}, \citenamefont {Bera}, \citenamefont {Lehtinen}, \citenamefont
  {Ronzani}, \citenamefont {Kohop\"a\"a}, \citenamefont {H\"onigl-Decrinis},
  \citenamefont {Shaikhaidarov}, \citenamefont {de~Graaf}, \citenamefont
  {Govenius},\ and\ \citenamefont {Prunnila}}]{mykkanen_enhancement_2020}%
  \BibitemOpen
  \bibfield  {author} {\bibinfo {author} {\bibfnamefont {E.}~\bibnamefont
  {Mykk\"anen}}, \bibinfo {author} {\bibfnamefont {A.}~\bibnamefont {Bera}},
  \bibinfo {author} {\bibfnamefont {J.~S.}\ \bibnamefont {Lehtinen}}, \bibinfo
  {author} {\bibfnamefont {A.}~\bibnamefont {Ronzani}}, \bibinfo {author}
  {\bibfnamefont {K.}~\bibnamefont {Kohop\"a\"a}}, \bibinfo {author}
  {\bibfnamefont {T.}~\bibnamefont {H\"onigl-Decrinis}}, \bibinfo {author}
  {\bibfnamefont {R.}~\bibnamefont {Shaikhaidarov}}, \bibinfo {author}
  {\bibfnamefont {S.~E.}\ \bibnamefont {de~Graaf}}, \bibinfo {author}
  {\bibfnamefont {J.}~\bibnamefont {Govenius}}, \ and\ \bibinfo {author}
  {\bibfnamefont {M.}~\bibnamefont {Prunnila}},\ }\bibfield  {title}
  {{\enquote {\bibinfo {title} {Enhancement of
  {Superconductivity} by {Amorphizing} {Molybdenum} {Silicide} {Films} {Using}
  a {Focused} {Ion} {Beam}},}\ }}\href {\doibase 10.3390/nano10050950}
  {\bibfield  {journal} {\bibinfo  {journal} {Nanomaterials}\ }\textbf
  {\bibinfo {volume} {10}},\ \bibinfo {pages} {950} (\bibinfo {year}
  {2020})}\BibitemShut {NoStop}%
\bibitem [{\citenamefont {Linzen}\ \emph {et~al.}(2017)\citenamefont {Linzen},
  \citenamefont {Ziegler}, \citenamefont {Astafiev}, \citenamefont {Schmelz},
  \citenamefont {H\"ubner}, \citenamefont {Diegel}, \citenamefont {Il{'}ichev},\
  and\ \citenamefont {Meyer}}]{linzen_structural_2017}%
  \BibitemOpen
  \bibfield  {author} {\bibinfo {author} {\bibfnamefont {S.}~\bibnamefont
  {Linzen}}, \bibinfo {author} {\bibfnamefont {M.}~\bibnamefont {Ziegler}},
  \bibinfo {author} {\bibfnamefont {O.~V.}\ \bibnamefont {Astafiev}}, \bibinfo
  {author} {\bibfnamefont {M.}~\bibnamefont {Schmelz}}, \bibinfo {author}
  {\bibfnamefont {U.}~\bibnamefont {H\"ubner}}, \bibinfo {author} {\bibfnamefont
  {M.}~\bibnamefont {Diegel}}, \bibinfo {author} {\bibfnamefont
  {E.}~\bibnamefont {Il{'}ichev}}, \ and\ \bibinfo {author} {\bibfnamefont
  {H.-G.}\ \bibnamefont {Meyer}},\ }\bibfield  {title} {{\enquote {\bibinfo {title} {Structural and electrical properties of
  ultrathin niobium nitride films grown by atomic layer deposition},}\ }}\href
  {\doibase 10.1088/1361-6668/aa572a} {\bibfield  {journal} {\bibinfo
  {journal} {Superconductor Science and Technology}\ }\textbf {\bibinfo
  {volume} {30}},\ \bibinfo {pages} {035010} (\bibinfo {year}
  {2017})}\BibitemShut {NoStop}%
\bibitem [{\citenamefont {Liang}\ and\ \citenamefont
  {Chen}(1996)}]{liang_interfacial_1996}%
  \BibitemOpen
  \bibfield  {author} {\bibinfo {author} {\bibfnamefont {J.~M.}\ \bibnamefont
  {Liang}}\ and\ \bibinfo {author} {\bibfnamefont {L.~J.}\ \bibnamefont
  {Chen}},\ }\bibfield  {title} {\enquote {\bibinfo {title} {Interfacial
  reactions and thermal stability of ultrahigh vacuum deposited multilayered
  {Mo}/{Si} structures},}\ }\href {\doibase 10.1063/1.361835} {\bibfield
  {journal} {\bibinfo  {journal} {Journal of Applied Physics}\ }\textbf
  {\bibinfo {volume} {79}},\ \bibinfo {pages} {4072--4077} (\bibinfo {year}
  {1996})}\BibitemShut {NoStop}%
\bibitem [{\citenamefont {Levy-Bertrand}\ \emph {et~al.}(2019)\citenamefont
  {Levy-Bertrand}, \citenamefont {Klein}, \citenamefont {Grenet}, \citenamefont
  {Dupr\'e}, \citenamefont {Beno\^it}, \citenamefont {Bideaud}, \citenamefont
  {Bourrion}, \citenamefont {Calvo}, \citenamefont {Catalano}, \citenamefont
  {Gomez}, \citenamefont {Goupy}, \citenamefont {Gr\"unhaupt}, \citenamefont
  {Luepke}, \citenamefont {Maleeva}, \citenamefont {Valenti}, \citenamefont
  {Pop},\ and\ \citenamefont
  {Monfardini}}]{levy-bertrand_electrodynamics_2019}%
  \BibitemOpen
  \bibfield  {author} {\bibinfo {author} {\bibfnamefont {F.}~\bibnamefont
  {Levy-Bertrand}}, \bibinfo {author} {\bibfnamefont {T.}~\bibnamefont
  {Klein}}, \bibinfo {author} {\bibfnamefont {T.}~\bibnamefont {Grenet}},
  \bibinfo {author} {\bibfnamefont {O.}~\bibnamefont {Dupr\'e}}, \bibinfo
  {author} {\bibfnamefont {A.}~\bibnamefont {Benoît}}, \bibinfo {author}
  {\bibfnamefont {A.}~\bibnamefont {Bideaud}}, \bibinfo {author} {\bibfnamefont
  {O.}~\bibnamefont {Bourrion}}, \bibinfo {author} {\bibfnamefont
  {M.}~\bibnamefont {Calvo}}, \bibinfo {author} {\bibfnamefont
  {A.}~\bibnamefont {Catalano}}, \bibinfo {author} {\bibfnamefont
  {A.}~\bibnamefont {Gomez}}, \bibinfo {author} {\bibfnamefont
  {J.}~\bibnamefont {Goupy}}, \bibinfo {author} {\bibfnamefont
  {L.}~\bibnamefont {Gr\"unhaupt}}, \bibinfo {author} {\bibfnamefont {U.~v.}\
  \bibnamefont {Luepke}}, \bibinfo {author} {\bibfnamefont {N.}~\bibnamefont
  {Maleeva}}, \bibinfo {author} {\bibfnamefont {F.}~\bibnamefont {Valenti}},
  \bibinfo {author} {\bibfnamefont {I.~M.}\ \bibnamefont {Pop}}, \ and\
  \bibinfo {author} {\bibfnamefont {A.}~\bibnamefont {Monfardini}},\ }\bibfield
   {title} {\enquote {\bibinfo {title} {Electrodynamics of granular aluminum
  from superconductor to insulator: {Observation} of collective superconducting
  modes},}\ }\href {\doibase 10.1103/PhysRevB.99.094506} {\bibfield  {journal}
  {\bibinfo  {journal} {Physical Review B}\ }\textbf {\bibinfo {volume} {99}},\
  \bibinfo {pages} {094506} (\bibinfo {year} {2019})},\ \bibinfo {note}
  {publisher: American Physical Society}\BibitemShut {NoStop}%
\bibitem [{\citenamefont {Zarea}, \citenamefont {Ueki},\ and\ \citenamefont
  {Sauls}(2023)}]{zarea_effects_2023}%
  \BibitemOpen
  \bibfield  {author} {\bibinfo {author} {\bibfnamefont {M.}~\bibnamefont
  {Zarea}}, \bibinfo {author} {\bibfnamefont {H.}~\bibnamefont {Ueki}}, \ and\
  \bibinfo {author} {\bibfnamefont {J.~A.}\ \bibnamefont {Sauls}},\ }\bibfield
  {title} {\enquote {\bibinfo {title} {Effects of anisotropy and disorder on
  the superconducting properties of niobium},}\ }\href
  {https://www.frontiersin.org/articles/10.3389/fphy.2023.1269872} {\bibfield
  {journal} {\bibinfo  {journal} {Frontiers in Physics}\ }\textbf {\bibinfo
  {volume} {11}} (\bibinfo {year} {2023})}\BibitemShut {NoStop}%
\end{thebibliography}
\end{document}